\documentclass{optica-article}

\journal{opticajournal}

\articletype{Research Article}

\usepackage{lineno}

\usepackage{siunitx}
\usepackage{tabularx}
\usepackage{booktabs}
\usepackage{float}
\usepackage{mathtools}
\usepackage{algorithm}
\usepackage{algpseudocode}

\def\c{\mathbf{c}}

\def\e{\text{e}}
\def\d{\text{d}}
\def\r{\mathbf{r}}
\def\epl{\varepsilon}
\DeclareMathOperator*{\argmin}{argmin}

\begin{document}

\nolinenumbers

\title{Model-based temporal unmixing towards quantitative photo-switching optoacoustic tomography}

\author{Yan Liu,\authormark{1}
Jonathan Chuah,\authormark{1}
Yishu Huang,\authormark{2}
Andre C. Stiel,\authormark{2,3}
Michael Unser,\authormark{1, *}
and Jonathan Dong\authormark{1}}

\address{\authormark{1}Biomedical Imaging Group, \'Ecole polytechnique f\'ed\'erale de Lausanne, Station 17, 1015 Lausanne, Switzerland\\
\authormark{2}Institute of Biological and Medical Imaging, Helmholtz Zentrum M\"unchen, Neuherberg, Germany
\\
\authormark{3}Faculty of Biology and Pre-Clinical Medicine, University of Regensburg, Regensburg, Germany
}
\email{\authormark{*}michael.unser@epfl.ch} 

\begin{abstract*}
Optoacoustic (OA) imaging combined with reversibly photoswitchable proteins has emerged as a promising technology for the high-sensitivity and multiplexed imaging of cells in live tissues in preclinical research.
Through carefully-designed illumination schedules of ON and OFF laser pulses, the resulting OA signal is a multiplex of different reporter species and the background. 
We propose a model-based variational framework to computationally unmix and image different species of photo-switching reporters using optoacoustic tomography. 
It is based on a detailed mathematical description of the photo-switching mechanism, which models how relevant physical parameters such as the kinetic constants and light fluence impact the switching signal. 
We introduce an algorithm that operates on images, as opposed to traditional pixelwise approaches.
It takes the form of an iterative inversion combined with tailored $\ell_1$ and total-variation regularization to increase the robustness to noise and to improve the unmixing quality.
We show that our method is able to disentangle multiple spatially overlapping labels and to recover continuous maps of quantities of interest on controlled phantoms and mice experiments. 
\end{abstract*}
\section{Introduction}
Optoacoustic tomography (OAT), also known as photoacoustic tomography, is a powerful and versatile imaging technology.
It gives in vivo access to a range of scales, from organelles to whole-body small animals at penetration depths from a few micrometers to centimeters
\cite{wang_photoacoustic_2012, yao_sensitivity_2014, taruttis_advances_2015, wang_practical_2016, yao_multiscale_2016, vu_listening_2019, dean-ben_optoacoustic_2019, omar_optoacoustic_2019, yao_perspective_2021}.
It has been successfully used in a wide range of preclinical and clinical studies such as blood oxygenation \cite{ wang2006noninvasive, li_photoacoustic_2018, Tomaszewski2017}, skin-cancer screening \cite{stoffels2015metastatic, he_fast_2022}, and breast imaging \cite{ermilov2009laser, diot2017multispectral, li2015high}.
OAT exploits the thermoacoustic effect: when chromophores (e.g. blood hemoglobin) are illuminated by short laser pulses, the absorbed photon energy is converted into ultrasonic waves by thermoelastic expansion of the medium surrounding them \cite{wang_photoacoustic_2012}.
Therefore, OAT benefits at the same time from the excellent contrast of optical excitation and from the large penetration depth of ultrasound \cite{wang_practical_2016, ntziachristos_going_2010}.

OAT uses a rich variety of endogenous contrast agents that enable label-free imaging and exogenous agents, such as genetically encoded protein reporters \cite{brunker_photoacoustic_2017}, that further enhance its specificity\cite{luke2012biomedical, wang_practical_2016, wang2004photoacoustic}. 
Among these reporters, the reversibly switchable optoacoustic proteins (rsOAPs) have emerged as a promising choice.
They offer increased sensitivity and allow one to image small clusters of fewer than one thousand labeled cells in mice \cite{li_small_2018, yao2016}.
The rsOAPs have a special photo-physical property of undergoing a change in their absorption spectrum when illuminated with specific wavelengths \cite{stiel_high-contrast_2015, yao_multiscale_2016, li_small_2018, mishra_photocontrollable_2019, mishra_genetically_2021,  li_multiscale_2021}.
This switching is reversible and robust to photobleaching \cite{chee2018, mishra_multiplexed_2020}.
Since our scheme is multiplexed temporally, we avoid spectral coloring, a typical issue in spectral unmixing \cite{li_photoacoustic_2018, dean-ben_light_2015, tzoumas_unmixing_2014}.
Combined with tailored illumination schedules, it can greatly improve the optoacoustic specificity: the weak modulated signals generated by the photo-switching targets can be extracted from the strong constant signals from the tissue background \cite{stiel_high-contrast_2015, dean-ben_light_2015, yao_multiscale_2016, li_small_2018, mishra_multiplexed_2020, mark_dual-wavelength_2018, kasatkina_optogenetic_2022, chen_reversibly_2024}.

Several pixelwise techniques have been developed to analyze time series of the photo-switching optoacoustic images. 
Differential imaging has been among the first methods being proposed \cite{yao2016}. 
It is based on a subtraction between the fully switched OFF- and ON- images (i.e. the last frame of each cycle), but can only account for a single photo-switching protein species.
Recent studies have proposed methods to unmix several species with different photo-switching rates by analyzing the whole time series, based on for example exponential fitting and Fourier analysis \cite{mishra_multiplexed_2020}, a specific labeling method with pure and mixed species of two proteins \cite{li_small_2018}, or reporters with non-overlapping spectra \cite{chee2018}. 
Yet these methods cannot extend to more than two types of labels or they require strict non-overlapping constraints.
A precise reconstruction based on a detailed description of the underlying physics of photo-switching would be the most informative and robust to noise. 
It requires a good characterization of the generation of the switching OA signal which depends on the intrinsic kinetic constants of the proteins and the intensity of local light fluence.

In this paper, we propose a reconstruction framework for temporal unmixing of photo-switching OAT without the restriction on the number of protein species. 
The foundation of our method is a physical model that describes the temporal multiplexing of the OA signal during the photo-switching process.
Our unmixing method operates on entire images instead of pixelwise, which enables us to leverage powerful and flexible regularization techniques to improve the quality of the unmixing and to produce coherent intensity maps.
We show on both simulated and experimental phantoms that we can faithfully recover concentration maps of multiple overlapping protein species, after we have estimated the distribution of light fluence inside the sample via the diffusion equation. We further demonstrate on a mouse model that our method successfully unmixes three species based on minimal assumptions on the physical model.
This unmixing is successful despite a simplistic estimate of the fluence.

\section{Methods}\label{sec:methods}
\subsection{Principles of photo-switching OAT}
\begin{figure}[tb]
    \centering
    \includegraphics[width=\textwidth]{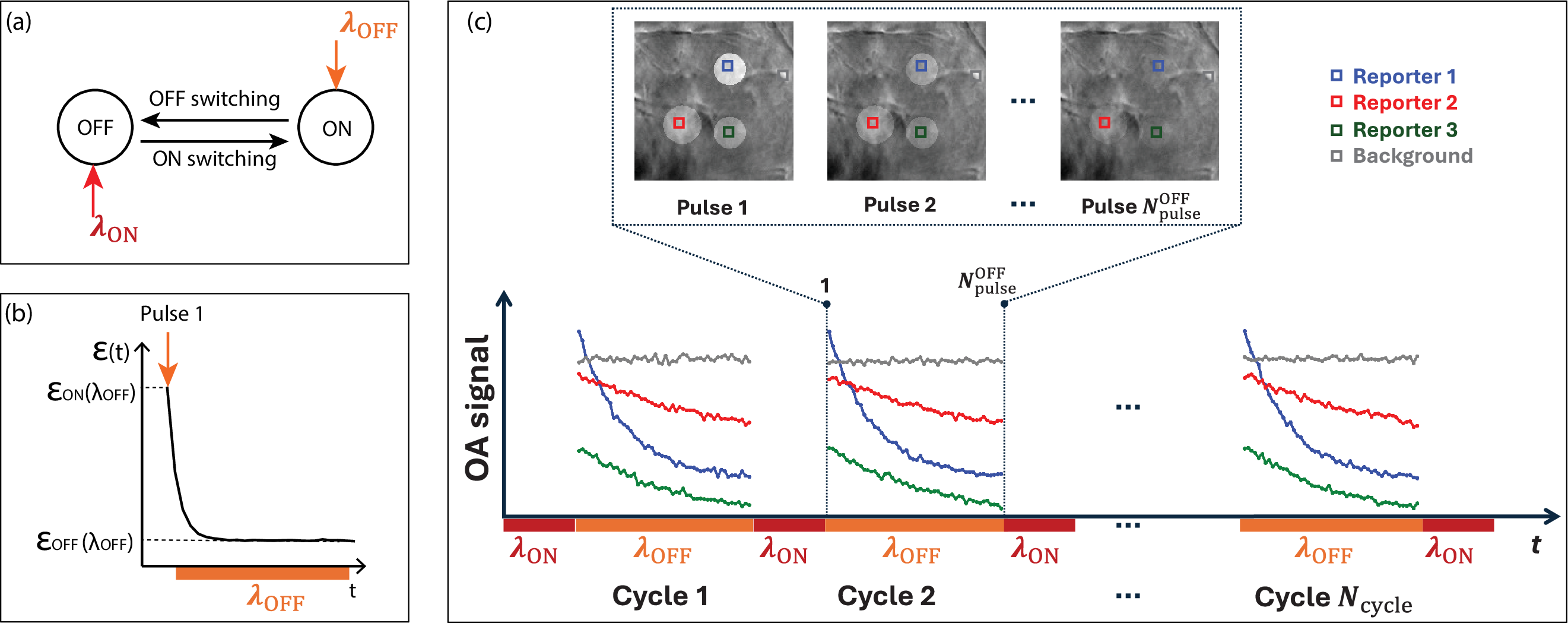}
    \caption{(a) Principle of reversible photo-switching OAT.
    (b) Illustration of the evolution of the molar extinction coefficient $\varepsilon$ of a photoswitchable reporter over time under the OFF-switching wavelength.
    (c) Top: Illustration of the OA images during OFF-switching. 
    Bottom: OA signal at the highlighted locations (three inside reporters in color and one in the background in gray). 
    The ON switching signals are omitted here.}
    \label{fig:principle}
\end{figure}
During a photo-switching experiment, $N^{\textrm{ON}}_\textrm{pulse}$ laser pulses at wavelength $\lambda_{\textrm{ON}}$ are first sent onto the sample until the proteins are switched to their ON state. 
Then, a sequence of $N^{\textrm{OFF}}_{\textrm{pulse}}$ pulses at the OFF-switching wavelength $\lambda_{\textrm{OFF}}$ progressively switches the proteins to their OFF state.
This sequence results in a decay of the OA signal from the photo-switching reporters that is well described by an exponential  model. 
This alternation of ON and OFF cycles, also called a switching cycle, is repeated $N_{\textrm{cycle}}$ times.
The background signal remains constant during this modulation process.
The principle of photo-switching OAT is summarized in Fig. \ref{fig:principle}.

Both the ON- and OFF- switching series can be used for unmixing.
However, the OFF-switching series are often preferred because they are slower, red-shifted, and have a higher dynamic range for all the photoswitchable proteins used so far \cite{stankevych_practical_2021}.
These properties simplify the analysis and allow for a larger imaging depth. 

\subsection{Mathematical Model of photo-switching and Unmixing}
\subsubsection{Forward Model in the Continuum}
We consider the general case of $P$ species of reporters distributed along the concentration map $c_p(\mathbf{r})$ ([$\mu\text{M}$]) for $p = 1, \ldots, P$, characterized by an intrinsic kinetic constant $k_p$ ([$\textrm{s}^{-1} \text{mJ}^{-1} \text{cm}^2$]). 
Note that we model the non-switching background as the last ($P$th) ``reporter'' with kinetic constant $k_P=0$.

Without loss of generality, we look at one OFF-switching pulse sequence and assume that all reporters are initially in their ON-state. 
The transition of the proteins from ON to OFF states is a stochastic process under the pulsed illumination of the OFF-wavelength,
which results in an exponential decay of the extinction coefficient $\varepsilon_p(\mathbf{r}, t; \lambda_{\text{OFF}})$ ([$\mu\textrm{M}^{-1}\textrm{cm}^{-1}$]):
\begin{equation}
    \label{eq:epsilon}
    \varepsilon_p(\mathbf{r}, t; \lambda_{\text{OFF}}) =
    \left(\varepsilon_p^{\text{ON}}(\lambda_{\text{OFF}}) - \varepsilon_p^{\text{OFF}}(\lambda_{\text{OFF}})\right)
    \text{e}^{-k_p\Phi(\mathbf{r})t} + \varepsilon_p^{\text{OFF}}(\lambda_{\text{OFF}}),
\end{equation}
where we adopt the approximation by Yao et al. \cite{yao_photoimprint_2014} that the decay rate depends linearly on the fluence $\Phi(\mathbf{r})$ ([$\textrm{mJ}/\textrm{cm}^2$]).
The distribution $H(\mathbf{r}, t)$ ([$\textrm{mJ}/\textrm{cm}^3$]) of total deposited optical energy at spatial location $\mathbf{r}$ and time $t$ is
\begin{equation}\label{eq:absorbed-energy-H}
    H(\mathbf{r}, t) = \Phi(\mathbf{r}) \sum_{p=1}^{P}\varepsilon_p(\mathbf{r}, t; \lambda_{\text{OFF}}) c_p(\mathbf{r}),\qquad \mathbf{r}\in\Omega, \quad t\ge0,
\end{equation}
where $\Omega$ represents the 3D sample to be imaged. 
We assume that the light pulses are independent from each other and that there is no temporal overlap with neighboring switching events.
The combination of Eq. \eqref{eq:epsilon} and Eq. \eqref{eq:absorbed-energy-H} yields the forward model
\begin{equation}\label{eq:final-H}
    H(\mathbf{r}, t) = 
    \Phi(\mathbf{r})\sum_{p=1}^{P}
    \left(\Delta\varepsilon_p
    \e^{-k_p\Phi(\mathbf{r})t} +
    \varepsilon_p^{\text{OFF}}\right)
    c_p(\mathbf{r}),
\end{equation}
where $\Delta\varepsilon_p =
\left(\varepsilon_p^{\text{ON}}(\lambda_{\text{OFF}}) -
\varepsilon_p^{\text{OFF}}(\lambda_{\text{OFF}})\right)$ and we omit the dependence of $\Delta\varepsilon_p$ and $\varepsilon_p^{\text{OFF}}$ on $\lambda_{\text{OFF}}$ to simplify notations.
In Eq. \eqref{eq:final-H}, the series of optical-energy deposition maps $H$ is proportional to $p^0$ ([mPa = mJ/$\textrm{cm}^3$]), the acoustically reconstructed OA image at the corresponding pulse number.
The unknown quantities are the concentration maps $c_p(\mathbf{r})$ of the $P$ species of reporters and the distribution $\Phi(\mathbf{r})$ of the light fluence within the sample.

The spatial distribution of fluence $\Phi(\mathbf{r})$ in the tomographic setting can be estimated separately via the diffusion equation and the optical parameters of the nonswitching background of the sample \cite{wang_biomedical_2007, bauer_quantitative_2011, cox_quantitative_2012}.
It is time-independent as the contribution of blood vessels to the absorption is much stronger than that of the protein reporters.
More details on the computation of fluence are provided in Section 2 of the Supplement.

\subsubsection{Discrete Forward Model}

To numerically implement the forward model, we discretize Eq. \eqref{eq:final-H} by sampling it over a compact domain $\Omega \subset \mathbb{R}^3$.
Because of tomographic sectioning, we consider a 2D cross-section of the sample and write the forward model in 2D while we still compute the light fluence in 3D.
We sample $H(\mathbf{r}, t)$ at spatial positions $\mathbf{r}_{i, j}$ for $i=0,\ldots, (L_{\text{x}} - 1)$, $j=0,\ldots, (L_{\text{y}} - 1)$, and time points (pulses) $t^{n}$ for $n=0, \ldots, (N-1)$. 
We define the energy deposition $h_{ij}^{n}= h(\mathbf{r}_{i, j},t^{n} )$, reporter concentrations $c_{p,ij} = c_{p} (\mathbf{r}_{i,j} )$, and fluence distribution $\hat{\Phi}_{ij} = \hat{\Phi} (\mathbf{r}_{i,j} )$.

The discrete linear system is first defined at a pixel $(i,j)$ for all time points as
\begin{equation}
    \label{eq:Sij}
    \underbrace{
    \begin{bmatrix}
    h^{0}_{ij} \\ h^{1}_{ij} \\ \vdots \\ h^{N-1}_{ij} 
    \end{bmatrix}}
    _{\mathbf{H}_{i,j}}
    = 
    \underbrace{
    \hat\Phi_{ijl}
    \begin{bmatrix}
    \Delta\varepsilon_{1} \e^{-k_{1} \hat{\Phi}_{ij} t^{0}} + \varepsilon_{1}^{\text{OFF}}
    &  
    \cdots & 
    \Delta\varepsilon_{P} \e^{-k_{P} \hat{\Phi}_{ij} t^{0}} + \varepsilon_{P}^{\text{OFF}}\\
    \Delta\varepsilon_{1} \e^{-k_{1} \hat{\Phi}_{ij} t^{1}} + \varepsilon_{1}^{\text{OFF}}
    & 
    \cdots & 
    \Delta\varepsilon_{P} \e^{-k_{P} \hat{\Phi}_{ij} t^{1}} + \varepsilon_{P}^{\text{OFF}}\\
    \vdots &  
    \ddots & 
    \vdots \\
    \Delta\varepsilon_{1} \e^{-k_{1} \hat{\Phi}_{ij} t^{N-1}} + \varepsilon_{1}^{\text{OFF}}
    & 
    \cdots & 
    \Delta\varepsilon_{P} \e^{-k_{P} \hat{\Phi}_{ij} t^{N-1}} + \varepsilon_{P}^{\text{OFF}}
    \end{bmatrix}}
    _{\mathbf{S}_{i, j}}
    \underbrace{
    \begin{bmatrix}
    c_{1,ij} \\ c_{2,ij} \\ \vdots \\ c_{P,ij} 
    \end{bmatrix}}
    _{\mathbf{C}_{i,j}},
\end{equation}
where $\mathbf{H}_{i,j} \in \mathbb{R}^{N}$, $\mathbf{C}_{i,j} \in \mathbb{R}^{P}$, and $\mathbf{S}_{i,j} \in \mathbb{R}^{N \times P}$. 
Then, we assemble the per-pixel systems of Eq. \eqref{eq:Sij} into a block-diagonal system by sequentially combining $L=L_{\text{x}} L_{\text{y}}$ systems 
\begin{equation}
    \label{eq:lin_model_blk}
    \underbrace{
    \begin{bmatrix}
    \mathbf{H}_{0,0} \\ \mathbf{H}_{1,0} \\ \vdots \\
    \mathbf{H}_{L_{\text{x}}-1,0} \\ \mathbf{H}_{0,1} \\ \vdots \\
    \mathbf{H}_{L_{\text{x}}-1,L_{\text{y}}-1}
    \end{bmatrix}}
    _{\mathbf{H}}
    = 
    \underbrace{
    \begin{bmatrix}
    \mathbf{S}_{0,0} & 
    \mathbf{0} &     
    \cdots & 
    \cdots & 
    \cdots & 
    \cdots & 
    \mathbf{0} \\
    \mathbf{0} & 
    \mathbf{S}_{1,0} &
    \ddots &     
    \cdots & 
    \cdots & 
    \cdots & 
    \vdots \\
    \vdots & 
    \ddots &
    \ddots &     
    \mathbf{0} & 
    \cdots & 
    \cdots & 
    \vdots \\
    \vdots & 
    \cdots & 
    \mathbf{0} &     
    \mathbf{S}_{L_{\text{x}}-1,0} &
    \mathbf{0} & 
    \cdots & 
    \vdots \\
    \vdots & 
    \cdots & 
    \cdots & 
    \mathbf{0} &     
    \mathbf{S}_{0,1} &
    \ddots & 
    \vdots \\
    \vdots & 
    \cdots & 
    \cdots & 
    \cdots & 
    \ddots &     
    \ddots &
    \mathbf{0} \\
    \mathbf{0} & 
    \cdots & 
    \cdots & 
    \cdots & 
    \cdots & 
    \mathbf{0} &     
    \mathbf{S}_{L_{\text{x}}-1, L_{\text{y}}-1} \\
    \end{bmatrix}}
    _{\mathbf{S}}
    \underbrace{
    \begin{bmatrix}
    \mathbf{C}_{0,0} \\ \mathbf{C}_{1,0} \\ \vdots \\
    \mathbf{C}_{L_{\text{x}}-1,0} \\ \mathbf{C}_{0,1} \\ \vdots \\
    \mathbf{C}_{L_{\text{x}}-1,L_{\text{y}}-1}
    \end{bmatrix}}
    _{\mathbf{C}},
\end{equation}
where the vector of measurement $\mathbf{H} \in \mathbb{R}^{NL}$, concentration maps $\mathbf{C} \in \mathbb{R}^{PL}$, the system matrix $\mathbf{S} \in \mathbb{R}^{NL \times PL}$.

By assembling the temporal signals of all pixels into one linear system, we aggregate all the available information to reconstruct concentration maps of all reporters at once.
Compared to pixelwise fitting, our approach enables spatial regularization to increase noise robustness and obtain more coherent concentration maps. 

\subsubsection{Unmixing Algorithm}

The unmixing task is to reconstruct individual maps of the concentration of reporters from measurements of the optical deposition over time via the linear system Eq. \eqref{eq:lin_model_blk}. 
In a typical photo-switching OAT experiment, $N \sim 10^1$, $P \sim 10^{0}$, and $L \sim 10^6$, which yields a linear system of size $\sim 10^{13}$.
Direct inversion of such a large sparse linear system is computationally challenging and not robust against measurement noise. 
We thus adopt a model-based approach that retrieves the solution to Eq. \eqref{eq:lin_model_blk} by solving the following optimization problem
\begin{equation} 
    \label{eq:recon_opt}
    \hat{\mathbf{C}} = \argmin{J(\mathbf{C})} 
     = \argmin_{\mathbf{C} \geq \mathbf{0}}\left\{\frac{1}{2} 
        \lVert \mathbf{H} - \mathbf{SC} \rVert^{2}_{2}
    +  \mathcal{R}(\textbf{C}) \right\},
\end{equation}
which involves a data-fidelity term and a regularization $\mathcal{R}(\mathbf{C})$. 
The regularization term enforces prior information on the solution, which we define as the combination of the total-variation (TV) and $\ell_1$ regularizers given by
\begin{equation}\label{eq:regularization}
    \mathcal{R}(\textbf{C}) = \gamma^{\text{TV}} \sum_{p=1}^{P} |\mathbf{C}^p|_{\text{TV}} + \gamma^{\ell_1} \sum_{(i,j)}^L \left(\sum_{p=1}^{P} \|\mathbf{C}^p_{i,j}\|_{\ell_1}\right),
\end{equation}
where $\mathbf{C}^p_{i,j}$ is the concentration of the $p$th reporter at pixel $(i,j)$, nonnegative constants $\gamma^{\text{TV}}$ and $\gamma^{\ell_1}$ are the weights of the TV and $\ell_1$ regularizers.

We choose the anisotropic TV regularizer $|\mathbf{C}|_{\text{TV}} = \lVert \nabla_{\text{x}} \mathbf{C} \rVert_{\ell_1} 
+ \lVert \nabla_{\text{y}} \mathbf{C} \rVert_{\ell_1}
$ for computational efficiency.
TV regularization promotes sparse gradients and hence achieves piecewise-smooth reconstructions \cite{chambolle_algorithm_2004, unser2016representer}. 
The $\ell_1$ regularization minimizes cross-talk between different species.

The minimization problem in Eq. \eqref{eq:recon_opt} is a large-scale nonlinear (due to the non-negativity constraint and regularization) convex problem.
For computational speed, we use the limited-memory  Broyden–Fletcher–Goldfarb–Shanno (BFGS) algorithm with bound constraints (L-BFGS-B \cite{lbfgsb}).
The forward matrices $\mathbf{S}_{i,j}$ are constructed efficiently using Einstein summation and  assembled into the final system matrix $\mathbf{S}$ in the form of a linear operator.
Finally, the whole pipeline, including the forward operator and the inverse-problem solver, is implemented using \emph{PyLops}, an open-source Python library for large-scale optimization with linear operators \cite{RAVASI2020100361}.

\subsection{Sample Preparation and Data Acquisition}\label{sec:exp-description}
We used three rsOAPs, ReBphP-PCM (Re), DrBphP-PCM (Dr) and RpBphP1-PCM (Rp), in the experiments. 
The protein was expressed in E. coli BL21 (DE3) cells, induced with IPTG, and purified using immobilized metal-affinity chromatography followed by size-exclusion chromatography. 
The ON and OFF spectra of the protein samples were recorded using a spectrophotometer (Shimadzu UV-1800, Shimadzu Inc., Kyoto, Japan) after being illuminated the samples with 660 nm or 780 nm LEDs (Thorlabs) in a 10 mm-pathlength quartz cuvette. 
Key photo-physical properties, including the absorption coefficient and the switching halftime for both OFF and ON wavelengths, are provided in Table 1 of the Supplement.
\begin{figure}[t]
    \centering
    \includegraphics[width=1.0\linewidth]{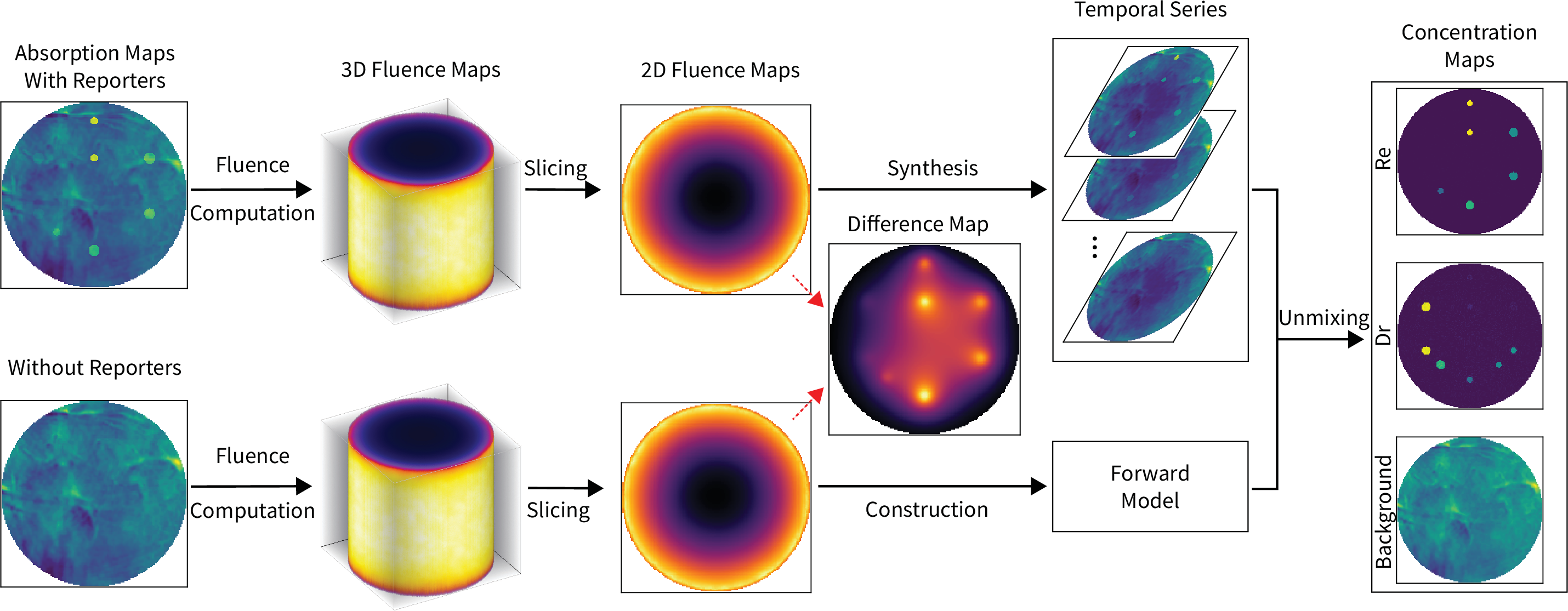}
    \caption{Workflow of the simulation pipeline.
    The absorption maps are 2D cross-sectional images of the original 3D maps.}
    \label{fig:simu-setup}
\end{figure}

Tissue-mimicking cylindrical phantoms (Section \ref{sec:exp1}) of diameter 27 mm were prepared by mixing 3\% intralipid (Sigma-Aldrich, St. Louis, Missouri, USA), 3\% sheep blood (Thermo Scientific, Massachusetts, USA), and 2\% agarose (Carl Roth, Karlsruhe, Germany) in a 50 ml flacon (30 mm diameter) as cast. 
The mixture aims to replicate the optical properties of biological tissues, with intralipid serving as scatterer, sheep blood as absorber, and agarose providing a gel matrix.
Protein samples were loaded into 580 $\mu$m catheter tubing and positioned at different locations within the phantom. 
We prepared two phantoms.
In one of them, we inserted four tubes with the following setting: one portion of Re (4.23 $\mu$M); one portion of Dr (3.71 $\mu$M); 1/3 of a portion of Re mixed with 2/3 of a portion of Dr; and 2/3 of a portion of Re mixed with 1/3 of a portion of Dr.
In the other phantom, we have three tubes, two of which contain one portion of Re and the last one one portion of Dr.
The two tubes of Re are placed at different depths from the boundary of the phantom, which results in a difference in local fluence.

Another phantom (Section \ref{sec:exp2}) is made of $1$ $\unit{\milli\meter}$-alginate beads inserted at random locations, 
each bead containing \emph{E. coli} that express one of the three rsOAPs. 
In the case of the mouse model, Jurkat T cells and \emph{E. coli} expressing each of the three rsOAPs are implanted into a 4T1 tumor on the back of a mouse in the same plane. 
We kindly refer the reader to \cite{mishra_multiplexed_2020} for details on sample preparation, data acquisition, and instrumental setup. 

Samples were scanned using MSOT InVision 256-TF (iThera Medical GmbH, Munich, Germany), described in more detail in \cite{mishra_multiplexed_2020} with a protocol of 50 pulses at 680 nm and 50 pulses at 770 nm.
The pulse train was repeated 50 times and an average was computed to reduce noise. 
The acquired data were reconstructed using the ViewMSOT software version 4.0.2.2 (iThera Medical GmbH, Munich, Germany) with the following settings: 100 $\mu$m resolution; 50 kHz to 6.5 MHz filtering; and a speed-of-sound trim of $(-30) \text{m}/\text{s}$.

\section{Results}\label{sec:results}

\subsection{Validation On Numerical Phantom}

We first validate our framework on a numerical phantom that is represented by a cylinder of diameter and height 27 mm.
The workflow of the simulation pipeline is illustrated in Fig. \ref{fig:simu-setup}.
The background medium is composed of an absorber (e.g., hemoglobin) with a uniform reduced scattering coefficient $\mu^{'}_s=10 \text{ cm}^{-1}$, a maximal absorption coefficient of $\mu_a=0.12 \text{ cm}^{-1}$ and a 2D absorption cross-section equal to the image corresponding to the last OFF-switching pulse from the experimental mouse dataset.
\begin{figure}[ht!]
    \centering
    \includegraphics[width=\textwidth]{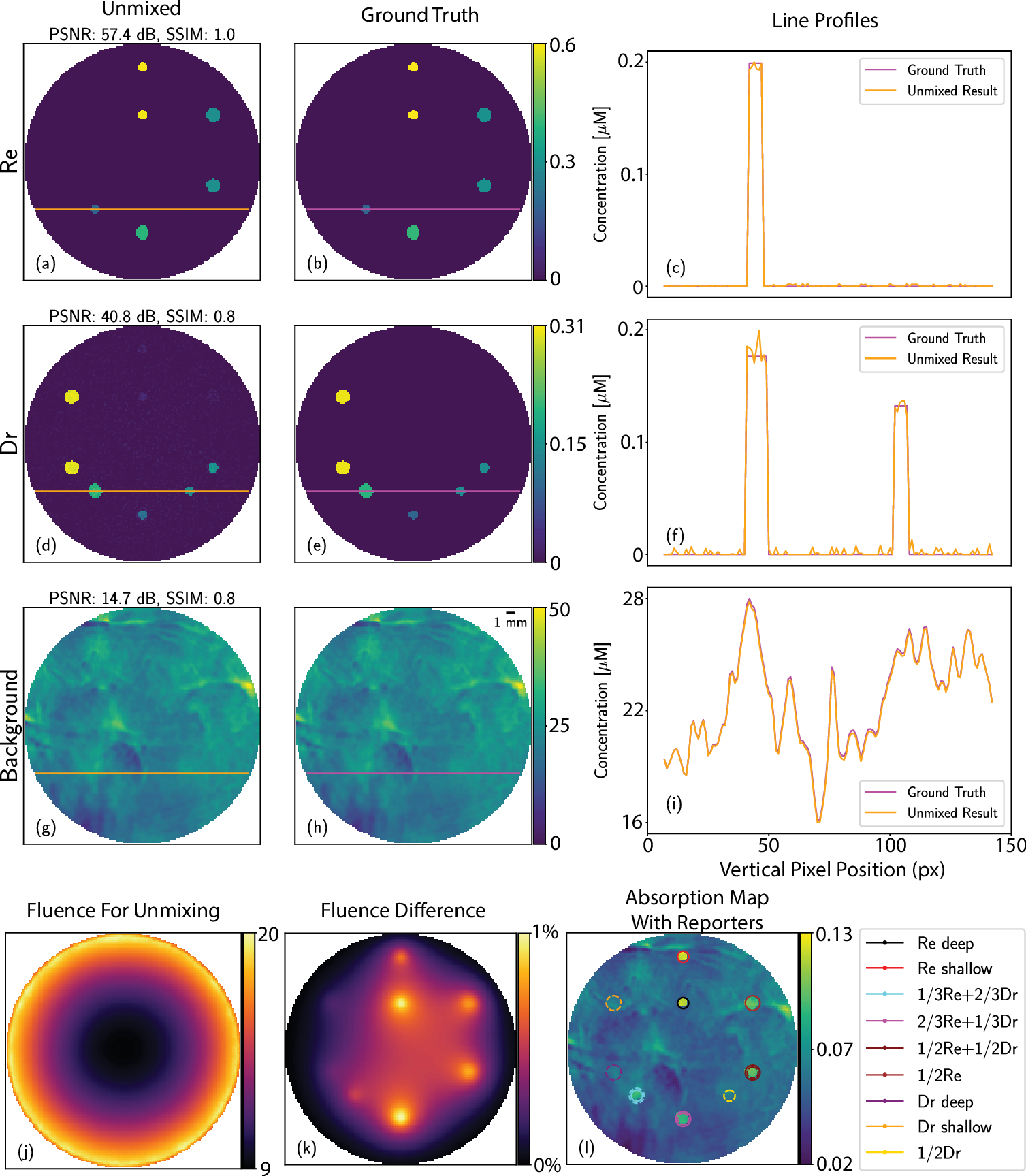}
    \caption{Unmixing results of the numerical simulation.
    (a), (d), (g): Unmixed quantitative maps of concentration in $\mu$M for Re, Dr, and the background, respectively.
    (b), (e), (h): Ground truth.
    The concentration maps of the unmixed and ground truth share the same scale (colorbar) for each reporter and the background.
    (c), (f), (i): Intensity profiles along a horizontal line across the concentration map (see for example (a) and (b)).
    (j) Fluence map used for the unmixing algorithm in arbitrary units.
    (k) Difference map of the fluence distribution with and without the contribution of photo-switching reporters, in percentage.
    (l) Absorption map with reporters.
    Solid and dashed circles represent the periphery of the Re and Dr species, respectively.
    The unit here is $\text{cm}^{-1}$.
    The ratios of the mixture of Re and Dr in each tube are indicated in the caption.
    }
    \label{fig:simu-phantom-results}
\end{figure}
We use a concentration of 0.6 $\mu$M of Re and 0.3 $\mu$M of Dr.
Then, we place nine tubes of two different radii in the background medium, 
with three tubes containing a mixture of Re and Dr according to different ratios, see Fig. \ref{fig:simu-phantom-results} (k) and (l) for details.
\begin{figure}[t]
    \centering
    \includegraphics[width=\textwidth]{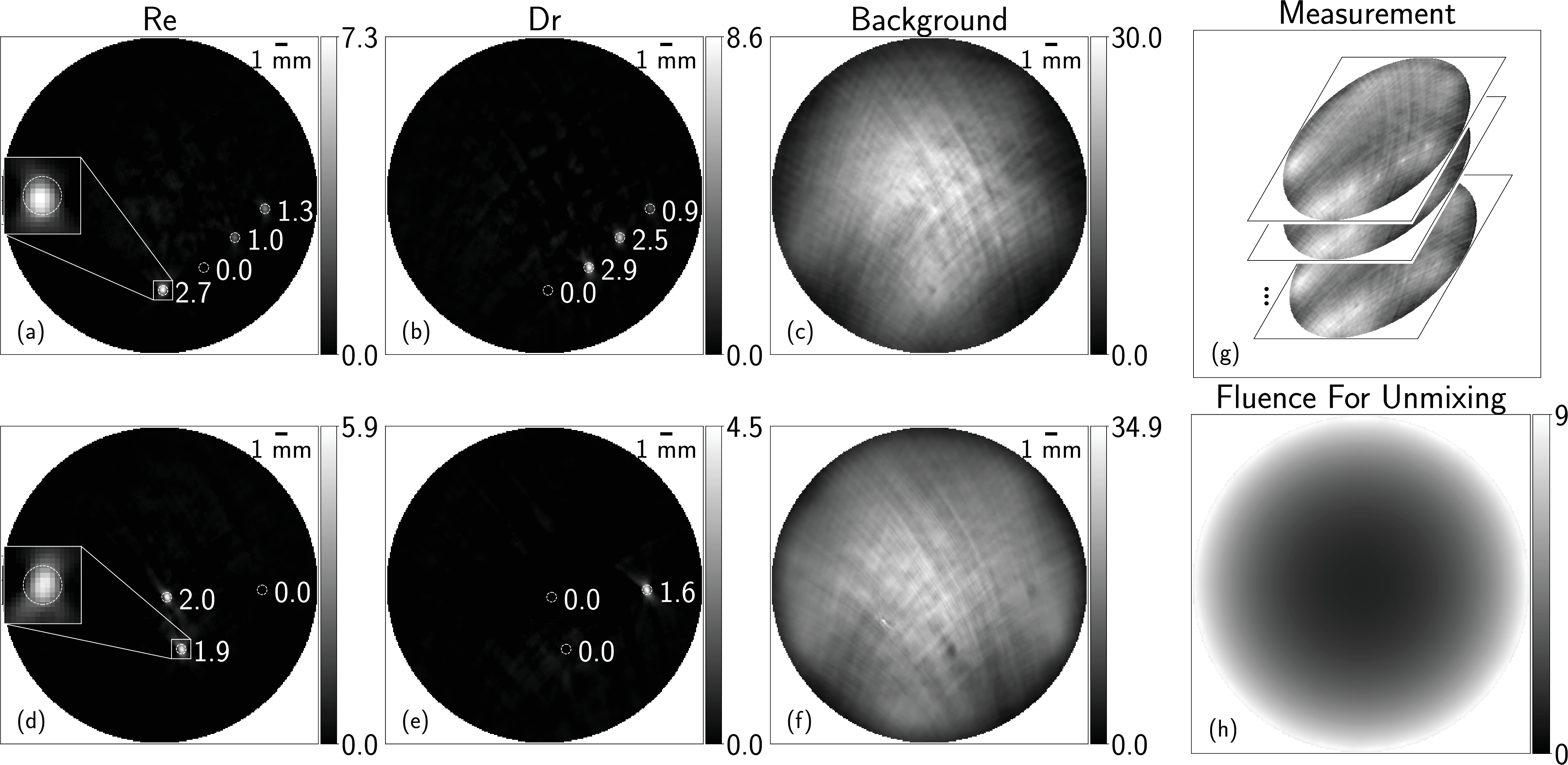}
    \caption{Unmixing results of tissue-mimicking phantoms.
    (a)-(c) Intensity maps of Re, Dr, and the background for the phantom with four tubes.
    (d)-(f) Intensity maps of Re, Dr, and the background for the phantom with three tubes.
    The ROI of each target is indicated by a dashed circle of radius 4 pixels.
    A close-up image of the ROI is provided for a Re target in (a) and (d).
    (g) Stack of three example images (first, second, and last frames) of OA measurements of the phantom with four tubes.
    (h) Fluence map used for the unmixing of both phantoms.
    All values are in arbitrary units.}
    \label{fig:phantom-results}
\end{figure}

We first compute the 3D fluence distribution via the diffusion equation described in Section 2 in the Supplement.
Then, we produce the time series of OA images following Eq. \ref{eq:lin_model_blk}.
Gaussian additive white noise of strength of 0.1\% of the maximum of the OA signal over the entire switching cycle is applied to the measurements.

To construct the forward model of unmixing, an estimation of the fluence distribution is required. 
As we assume no prior knowledge of the reporters to image, we compute the fluence using only the absorption map of the background. 
We show in Fig. \ref{fig:simu-phantom-results} (j) the fluence maps with the sole contribution of the background and additionally of the reporters in percentage.
The mismatch of these two maps is negligible (maximally 1\%) and thus will have little impact on the forward model and the unmixing quality.
We then perform unmixing by solving the nonlinear (due to the nonnegativity constraint) optimization problem defined in Eq. \ref{eq:recon_opt}.

We show in Fig. \ref{fig:simu-phantom-results} that
the quantitative maps of the concentration of both the reporters and the background are faithfully reconstructed.
Additional simulations that demonstrate the effectiveness of the regularization are provided in Section 3 of the Supplement.

\subsection{Experimental Results With Fluence Estimation}
\label{sec:exp1}

We then validate our framework on the tissue-mimicking blood phantoms described in Section \ref{sec:exp-description}. 
The workflow to estimate the fluence distribution and to construct the forward model for unmixing is similar to the simulation pipeline described in Fig. \ref{fig:simu-setup}, 
except that we assume a homogeneous absorption map of constant value $0.21 \text{ cm}^{-1}$ estimated based on the composition of the phantoms (see Section 2.B for details).
We show the fluence map and a series of OA measurements used for unmixing in Fig. \ref{fig:phantom-results} (g) and (h).

For the quantitative analysis, we define the intensity of a target as the sum of the intensities inside a circular region-of-interest (ROI) of radius 4 pixels around the brightest pixel inside the target divided by the number of pixels in the ROI.
We indicate the ROI with dotted circles and annotate the intensity of a target next to it in Fig. \ref{fig:phantom-results} (a), (b), (d), and (e).
We show that the unmixing of Re and Dr is successful for the phantom with three tubes in Fig. \ref{fig:phantom-results} (d) and (e).
Further, the ratio between the two tubes of Re that contain an equal amount of it is 2:1.9, which is very close to the true ratio of 1:1.
The result on the phantom that contains mixtures of Re and Dr shows a successful separation of these two species in all four tubes in Fig. \ref{fig:phantom-results} (a) and (b).
The ratio between the three tubes is 2.7:1.0:1.3 for Re and 2.9:2.5:0.9 for Dr, while the reference values are 3:1:2 and 3:2:1.
The mismatch between the recovered and the reference ratios arises from the imperfect illumination and heterogeneity of absorption within the phantom.
The code and data to reproduce the simulation and experimental results are made available online \cite{code_data}.
\subsection{Experimental Results With Three Species of Reporters}\label{sec:exp2}

We finally evaluate our approach on a phantom of beads and on a mouse model with three different types of reporters.
The experimental phantom datasets contain 2D OA images of size ($300\times300$) px of 50 OFF-switching pulses averaged over 50 OFF-switching cycles to reduce detection noise. 
Further, only the first 25 pulses of the experimental phantom datasets are used for unmixing as dynamics are dominated by noise fluctuations afterwards (see Fig. S1 of Supplement 1 for details).
The mouse dataset of 25 OFF-switching pulses averaged over 50 cycles contains 2D OA images that are cropped from ($332\times332$) px down to ($120\times90$) px to minimize the area devoid of switching information and thus reduce computational time.

\begin{figure}[tb]
    \centering
    \includegraphics[width=\textwidth]{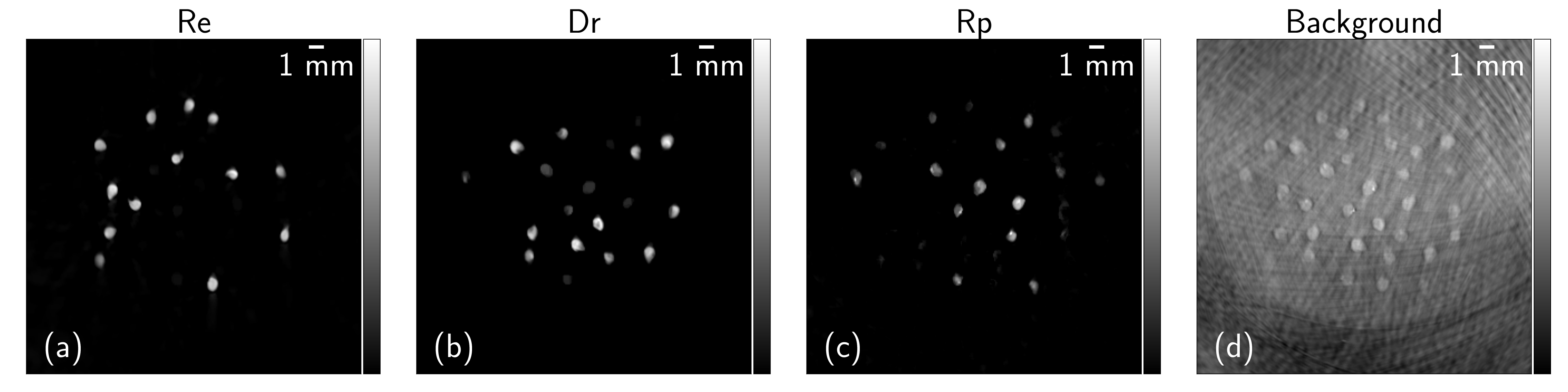}\\ 
    \includegraphics[width=\textwidth]{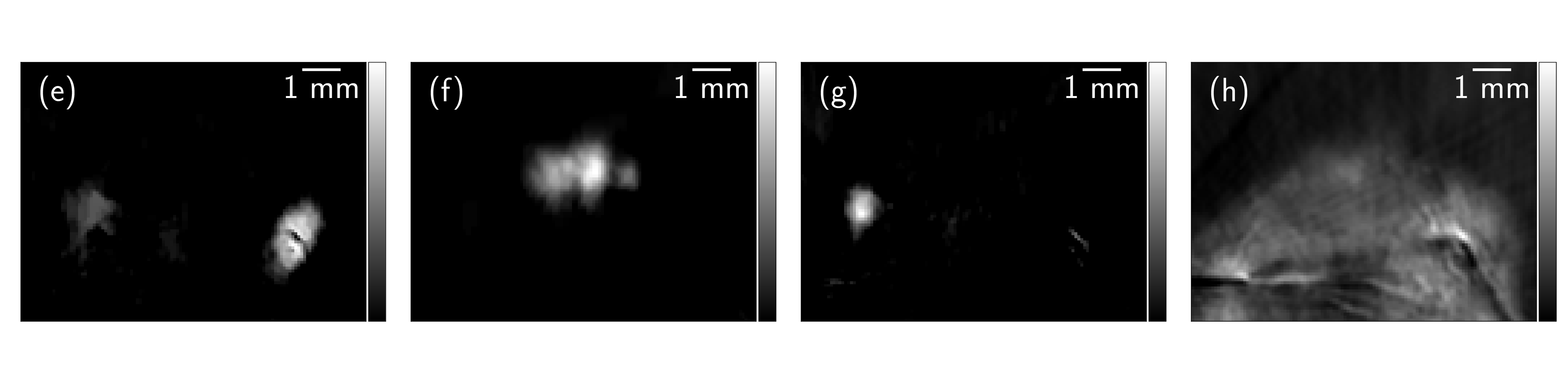}\\
    \includegraphics[height=0.24\textwidth]{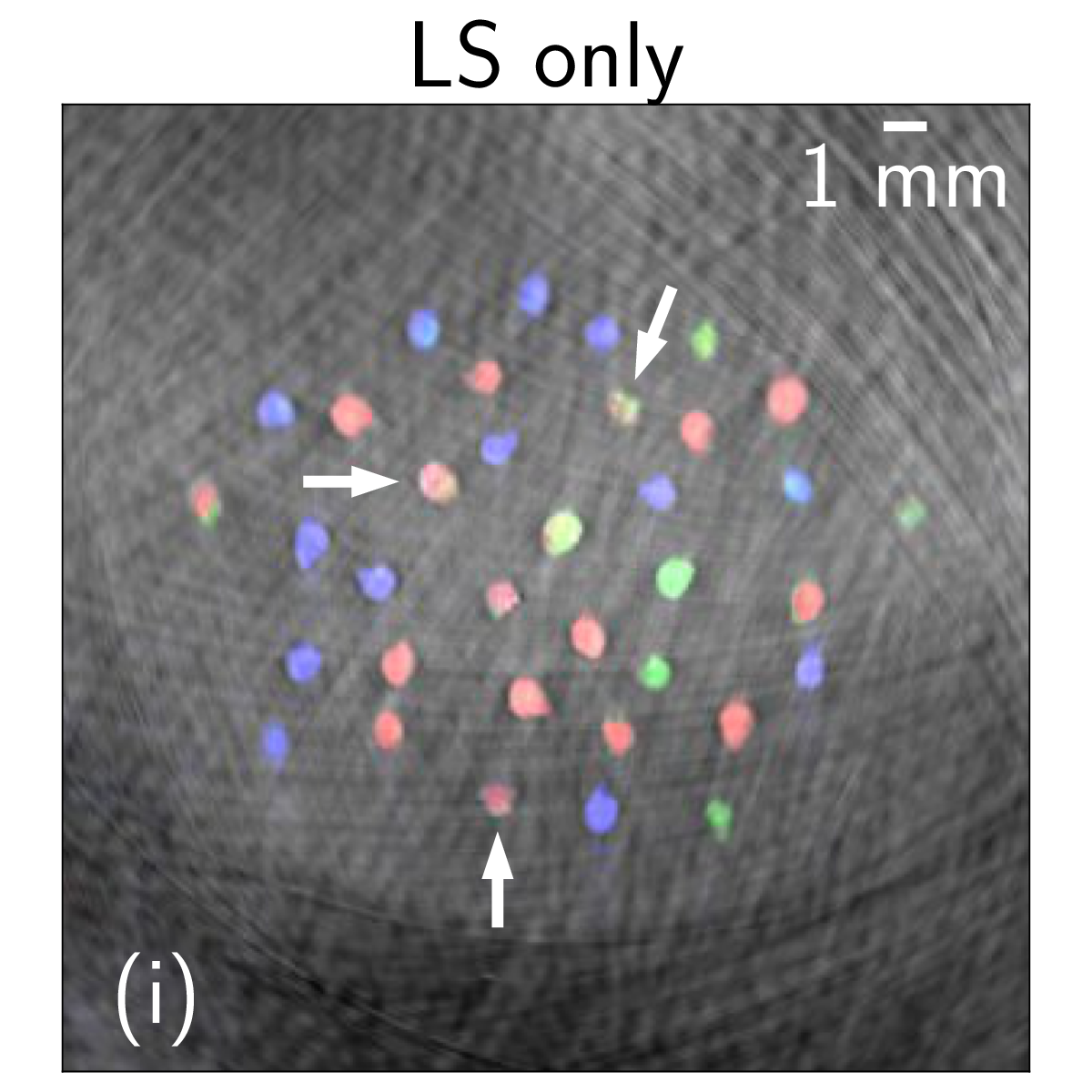}
    \includegraphics[height=0.24\textwidth]{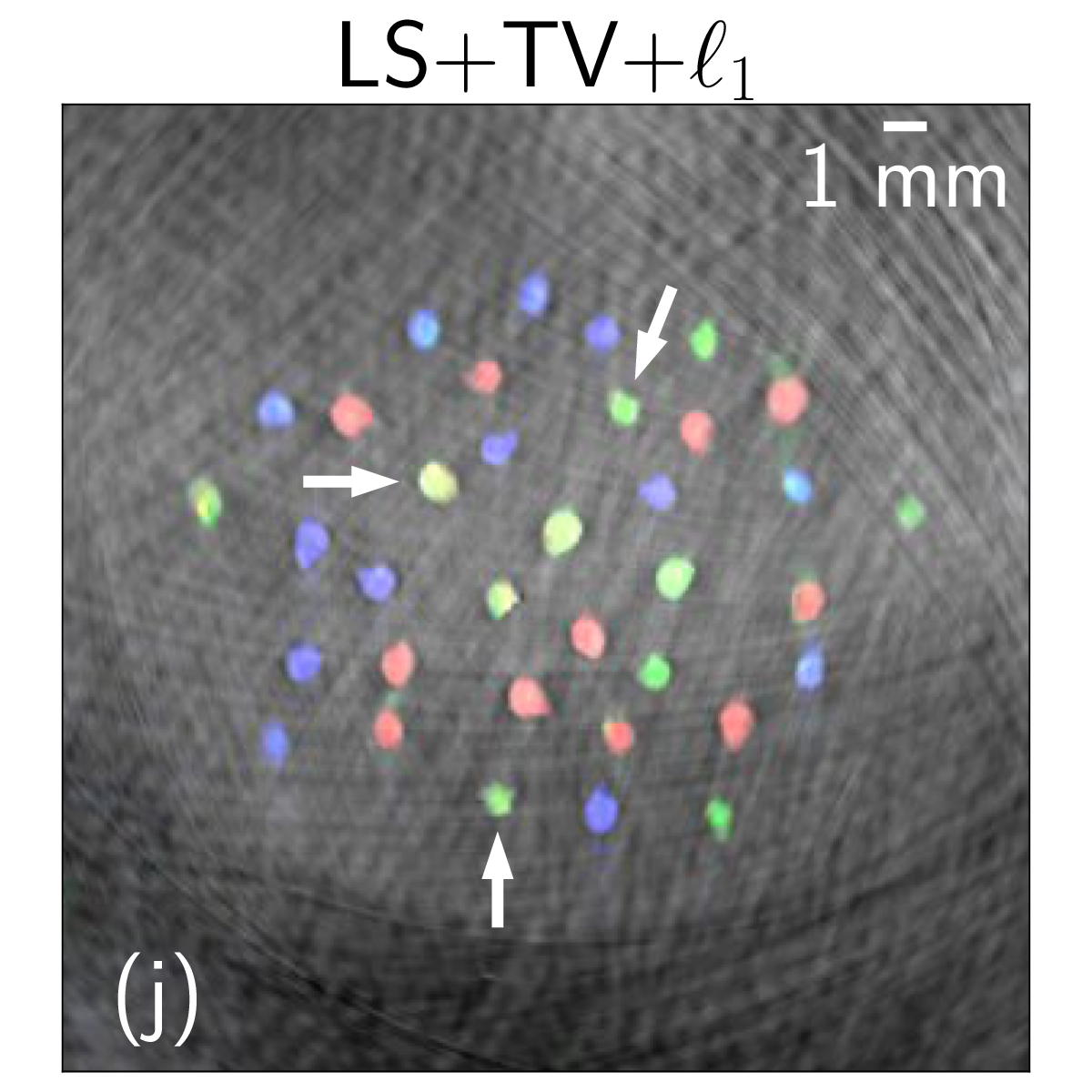}
    \includegraphics[height=0.24\textwidth]{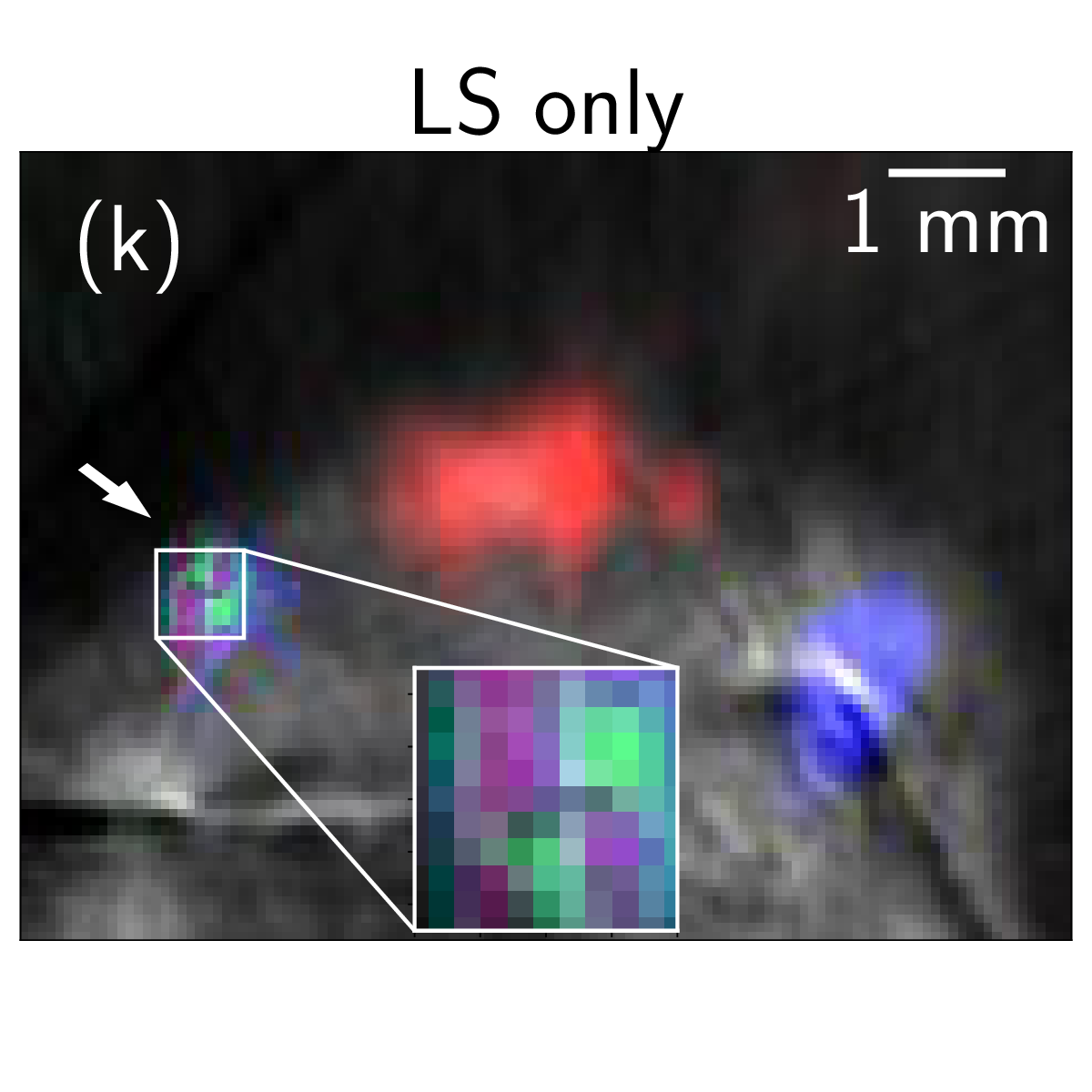}
    \includegraphics[height=0.24\textwidth]{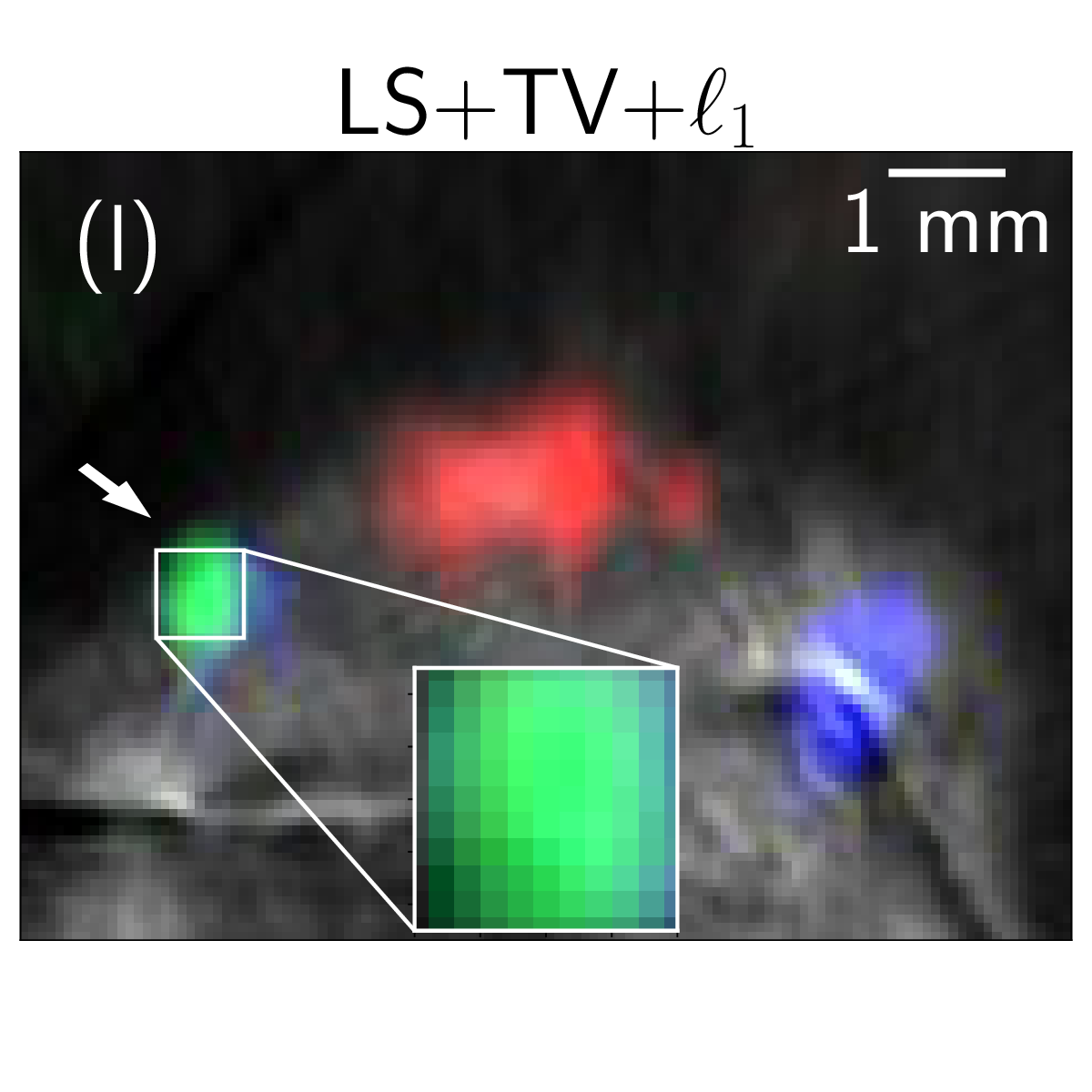}

    \caption{Unmixing results of a phantom of beads and mouse model.
    First two rows (a)-(h): reconstructed concentration maps of rsOAPs Re ((a) and (e)), Dr ((b) and (f)), and Rp ((c) and (g)) using the proposed combination of TV and $\ell_1$ regularization. (d) and (h): background.
    The color images (i)-(l) visualize the maps of the three reconstructed rsOAPs overlaid on the background (produced using Fiji \cite{schindelin_fiji_2012}). 
    Re, Dr and Rp are represented by blue, red, and green channels. The background is grayscale. 
    Color images (i) and (k) represent the reconstruction result without any regularization; (j) and (l) represent the result with our proposed regularization method.
    The arrows in (i) point out some of the beads belonging to the species Rp that were unsuccessfully unmixed under unregularized reconstruction compared to the successful outcome with regularization in (j).
    We provide a zoom-in of the details of the unmixing results of Rp in (k) and (l) to ease the  comparison of the unregularized and regularized methods.
    The intensities of all the images are normalized to [0, 1] for cross-comparison and color visualization.
    }
    \label{fig:experiments}
\end{figure}
To build the forward matrix $\mathbf{S}$ for the experimental dataset, the knowledge of the distribution of the light fluence and the intrinsic kinetic constants $k_p$ is required.
We adopt a constant fluence distribution $\Phi_0$ over space and time because of the tiny contribution of the reporter proteins compared to the tissue background and the shallow implantation of the proteins.
More details on the determination of the kinetic constants are presented in Section 2 of the Supplement.

The results in Fig. \ref{fig:experiments} (a)-(h) show that our unmixing algorithm successfully reconstructs the continuous spatial distribution of the three rsOAPs both for the beads phantom and for the mouse model.  
The overlay images of Fig. \ref{fig:experiments} allow us to compare the results of our proposed approach (Fig. \ref{fig:experiments} (j) and (l)) against those of the unregularized approach (Fig. \ref{fig:experiments} (i) and (k)).
We observe that, without any regularization, some beads belonging to the Rp channel show up in the Dr channel for instance those pointed out by the arrows in Fig. \ref{fig:experiments} (i) and (j).
The effect of regularization is even more obvious when we zoom in onto the Rp species in the reconstruction result of the mouse images in Fig. \ref{fig:experiments} (k) and (l).
In the absence of regularization, the Rp reporter cannot be properly unmixed from the other two, as indicated by the zoom-in region in Fig. \ref{fig:experiments} (k).
With our proposed regularization, the same region is cleanly unmixed with coherent intensity distribution,  as observed in Fig. \ref{fig:experiments} (l).

\section{Conclusion and Discussion}\label{sec:conclusion}
We presented a model-based unmixing framework to reconstruct the concentration maps of multiple species of photo-switching reporters using optoacoustic tomography.
This model describes how the time series of OA images evolve during the OFF-switching process.
Our model explicitly relates the unknown concentration maps to the factors that govern the switching dynamics, namely, the reporter kinetic constant, the extinction coefficient, and the distribution of light fluence in the sample. 
We make the best use of the full time-series information during OFF-switching cycles on the whole image, thus avoiding the shortcomings of pixelwise analysis.
This also enables the addition of carefully designed spatial regularization to the inverse reconstruction problem.
The combination of total variation and $\ell_1$ regularizers helps reduce noise and improves the separation of different species of reporters, which is crucial to a successful unmixing and high-quality reconstruction.

The proposed framework offers fast and efficient quantitative analysis of the temporally entangled OA data during photo-switching experiments.
This approach is versatile and easily applicable to various geometries such as 2D cross-sectioning or whole 3D volumes acquired with a raster-scanning or tomographic OA setup.
However, one limitation of our method is the assumption that the decay rate depends linearly on the fluence. This simplifying assumption is common in the field.
Our framework also allows for a realistic map of the heterogeneous light fluence distribution, which can be separately estimated with another model, to be conveniently incorporated in the forward model for high-quality volume unmixing.

\begin{backmatter}
\bmsection{Funding}
This project was funded by the European Union's Horizon Europe research and innovation programme under Grant Agreement No. 101046667 (SWOPT).

\bmsection{Acknowledgments}
We would like to thank Eric Sinner for his help in the code base.

\bmsection{Disclosures}
The authors declare no conflicts of interest.

\bmsection{Data availability} Phantom data underlying the results presented in this paper are available in \cite{code_data}. Mouse data are not publicly available at this time but may be obtained from the authors upon reasonable request.

\end{backmatter}

\bibliography{reference}

\newpage
\appendix

\section{Supporting Tables and Figures}\label{sec:figs}
We provide an extract of the key photo-physical properties of the reversibly switchable OA reporters  (rsOAPs) used in the experiments in Table \ref{tab:reporters}. 
For a summary of the complete properties of these reporters, refer to \cite{mishra_multiplexed_2020}. 
\begin{table}[h!]
    \centering
    \begin{tabular}{c|cccc}
    \toprule
        Name & 
        $\tau_{\text{OFF}}$  & $\bar{k}$ 
        & $\epl^{\text{ON}}_{\text{770nm}}$  & $\epl^{\text{OFF}}_{\text{770nm}}$ \\
        \midrule
        Unit & s  & $\text{s}^{-1}$ & $\mu\text{M}^{-1}\text{cm}^{-1}$ & $\mu\text{M}^{-1}\text{cm}^{-1}$ \\
        \midrule
        \emph{Re}BphP-PCM (Re) & 0.18 & 3.85 & $0.07$ & $0.002$ \\

        \emph{Dr}BphP-PCM (Dr) & 0.70 & 0.99 & $0.04$ & $0.004$ \\

        \emph{Rp}BphP1-PCM (Rp) & 0.42 & 1.65 & $0.06$ & $0.009$ \\
        \bottomrule
    \end{tabular}
    \caption{Key photo-physical properties of the three rsOAPs used in the experiments: half time of the OFF switching cycle $\tau_{\text{OFF}}$; kinetic constant $\bar{k}$ derived based on $\tau_{\text{OFF}}$; molar extinction coefficients of the ON and OFF states $\epl^{\text{ON}}_{\text{770nm}}$, $  \epl^{\text{OFF}}_{\text{770nm}}$ at  $770\unit{\nano\meter}$ (OFF-switching) wavelength.}
    \label{tab:reporters}
\end{table}

\begin{figure}[h!]
    \centering
    \includegraphics[width=0.8\textwidth]{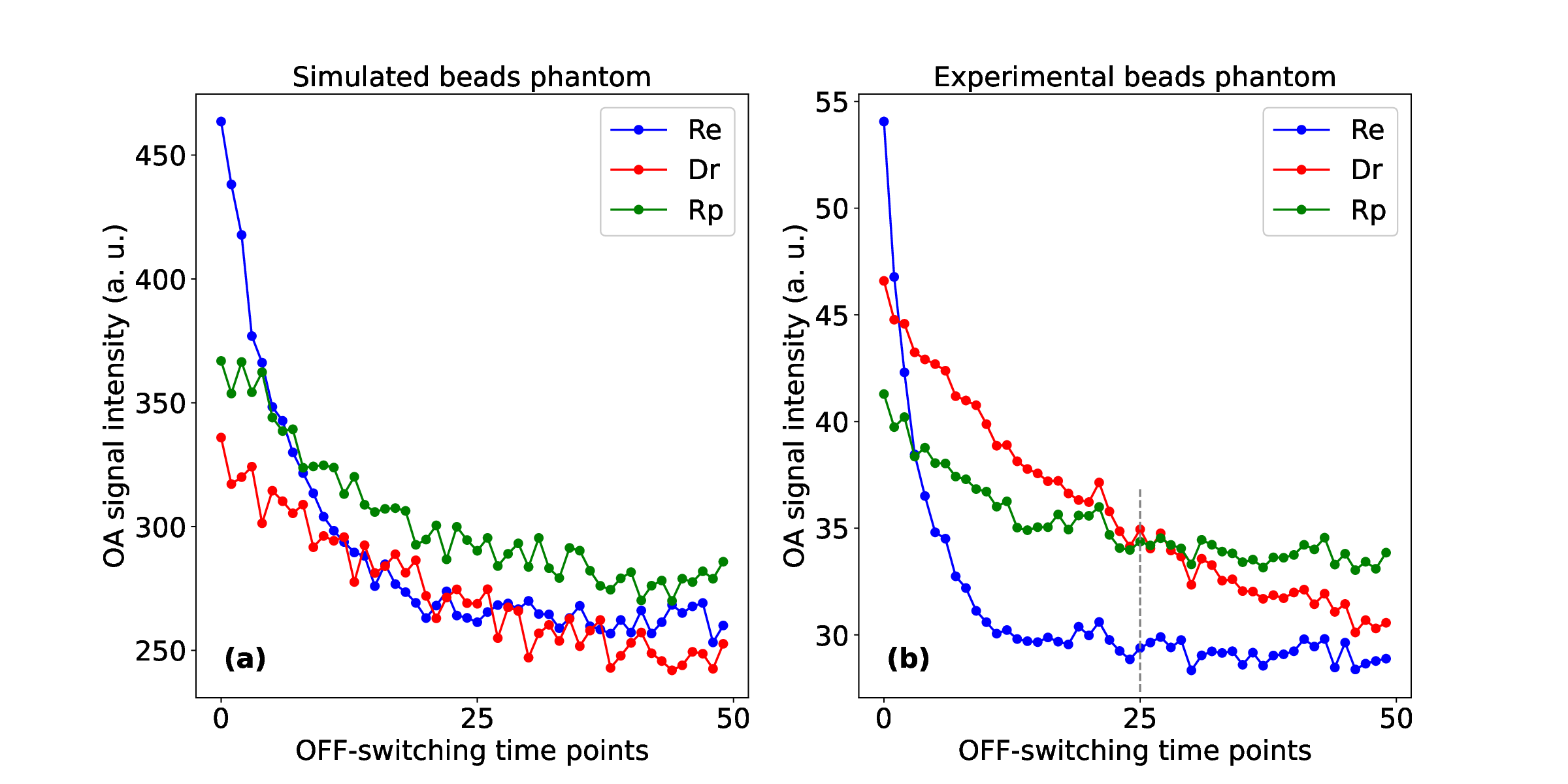}
    \caption{Evolution of the OA signal during an OFF-switching cycle of 50 time points at three representative locations inside the three rsOAPs: Re (blue), Dr (red), and Rp (green) for the simulated (a) and experimental measurements (b) of a phantom of beads.
    The noise level in the measurements of the simulated beads phantom is 1\% of its maximal intensity. }
    \label{fig:noise-compare}
\end{figure}

We provide in Fig. \ref{fig:noise-compare} the OA signal during a typical OFF switching cycle in the numerical simulation of a phantom of beads described in section 3 in this Supplement. 
We observe that after 25 time steps (marked with a vertical gray dashed line), there is less useful information in the signal other than noise.
Hence, only the first 25 frames of OA images are used in the reconstruction. 
\section{Estimation of the Light Fluence} \label{sec:fluence}
\subsection{Geometry of the Setup}
\begin{figure}[t]
    \centering
    \includegraphics[width=0.5\linewidth]{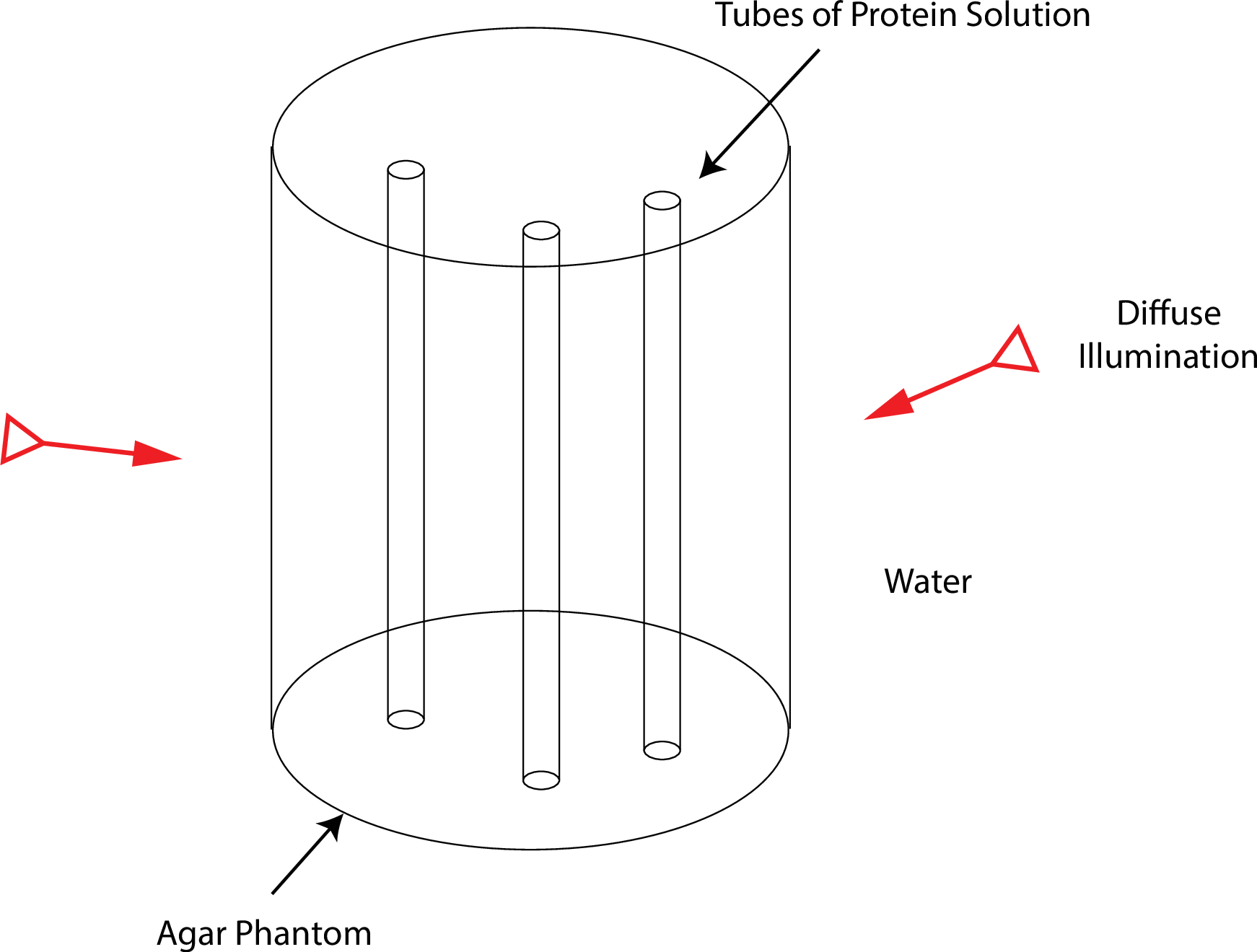}
    \caption{Experimental setup of the phantom.}
    \label{fig:phantom}
\end{figure}

The setup of our phantom experiment is illustrated in Fig. \ref{fig:phantom}.
The cylindrical agar phantom, which serves as the non-switching background, has a diameter of 27mm and is submerged in water to match the refractive indices.
We inserted tubes of diameter 580 $\mu$m filled with solutions of photo-switching proteins inside the phantom.
The illumination originates from five laser outlets that surround the body of the phantom such that the incident light can be assumed homogeneous and diffuse on the surface of the body of the cylinder.

\subsection{Optical Parameters}
The absorption coefficient $\mu_a$ of the background is calculated according to \cite{absorption}
\begin{equation}
    \label{eq:mu_a-blood}
    \mu_a(\lambda) = \ln 10 \cdot
    \varepsilon(\lambda) [\text{cm}^{-1 }\cdot \text{mole}^{-1} \cdot \text{L}] \cdot   \frac{x\% \cdot 150 [\text{g}\cdot\text{L}^{-1}]} 
    {64500 [\text{g}\cdot\text{mole}^{-1}]},
\end{equation}
where $\lambda=770$ nm is the OFF-switching wavelength, 
$\varepsilon(\lambda) = 1311.88 [\text{cm}^{-1 }\cdot \text{mole}^{-1} \cdot \text{L} = \text{cm}^{-1 }\cdot \text{M}^{-1}]$ is the molar extinction coefficient of blood at 770 nm, assuming the blood we used in the phantom to be mostly composed of deoxygenated hemoglobin (Hb), 
$x=3$ is the percentage of blood used in the phantom,
$150[\text{g}\cdot\text{L}^{-1}]$ is the typical mass concentration of hemoglobin within blood, 
and $64500 [\text{g}\cdot\text{mole}^{-1}]$ is the molecular weight of hemoglobin.
Using these values, the background is determined by $\mu_a^{\text{bg}} =  0.21 [\text{cm}^{-1}]$.

The absorption coefficients of the protein species Re and Dr are calculated via \cite{jacques_optical_2013}
\begin{equation}
    \label{eq:mu_a-protein}
    \mu_a(\lambda) = \ln 10 \cdot \varepsilon(\lambda) [\text{M}^{-1}\cdot\text{cm}^{-1}] \cdot c [\text{M}]
\end{equation}
from their molar concentration $c_{\text{Re}}=4.23\times 10^{-6}$ [M], 
$c_{\text{Dr}}=3.71\times 10^{-6}$ [M] and their respective molar extinction coefficients at 770nm $\varepsilon_{\text{Re}}(770\text{nm})\approx 0.7\times 10^{5}$ $[\text{M}^{-1}\cdot\text{cm}^{-1}]$, 
$\varepsilon_{\text{Dr}}(770\text{nm})\approx 0.4\times 10^{5}$ $[\text{M}^{-1}\cdot\text{cm}^{-1}]$.
We thus have $\mu_a^{\text{Re}} = 0.68\text{ cm}^{-1}$ and $\mu_a^{\text{Dr}} = 0.34\text{ cm}^{-1}$.

We adopt a typical value of $10\text{ cm}^{-1}$ for the reduced scattering coefficient $\mu_s^{'}$ of the background, which is composed of 3\% intralipid as a scatterer.

\subsection{Diffusion Equation}
From the calculated optical parameters, we see that the light propagation within the sample is within the diffuse regime in which scattering is much stronger than absorption ($\mu_a\ll \mu_s^{'}$).
We thus adopt the diffusion equation \eqref{eq:de} complemented by the suitable boundary condition \eqref{eq:bc} which describes the tissue/water interface to compute the light fluence 
distribution $\Phi(\mathbf{r})$ over a sample $\Omega$ as
\cite{wang_biomedical_2007, ammari_mathematical_2012}
\begin{equation}
    \label{eq:de}
    \mu_a(\mathbf{r})\Phi(\mathbf{r}) - \nabla\cdot (D(\mathbf{r})\nabla\Phi(\mathbf{r}))=S(\mathbf{r}), \qquad \mathbf{r}\in\Omega,
\end{equation}

\begin{equation}
    \label{eq:bc}
    \Phi(\mathbf{r}) - 2D(\mathbf{r})\nabla \Phi(\mathbf{r})\cdot\mathbf{n}=0, \qquad\mathbf{r}\in\partial\Omega
\end{equation}
where $\Omega$ is a 3D domain, and
$\mathbf{n}$ is the outward normal vector of the boundary.
The illumination is described by a function $S(\mathbf{r})$.
The diffusion coefficient $D(\mathbf{r})$ depends on the absorption coefficient $\mu_a(\mathbf{r})$ and the reduced scattering coefficient $\mu^{'}_s(\mathbf{r})$ according to 

\begin{equation}
    \label{eq:diffusion-coefficient}
    D(\mathbf{r})=\frac{1}{3(\mu_a(\mathbf{r}) + \mu^{'}_s(\mathbf{r}))}.
\end{equation}

In the case of our agar cylinder phantom, we model the domain $\Omega$ with a cylinder of height $H$ and diameter $L$ (see Fig. \ref{fig:relation-meshes}), which leads to

\begin{equation}
    \label{eq:domain}
    \Omega = \left\{(x, y, z) | (z - z_c)^2 + (x - x_c)^2 \leq (L/2)^2, 0\leq y - y_c \leq H\right\}.
\end{equation}
 
The incident diffuse illumination on the body of the phantom
is approximated by a function $S(\mathbf{r})$ of constant-intensity $I_0$ 

\begin{equation}
    \label{eq:laser-source}
S(x, y, z) = \left\{
\begin{aligned}
    & I_0, \text{ if }  (z - z_c)^2 + (x - x_c)^2 = (L/2)^2 \\ 
    & 0, \text{ else}.
\end{aligned}
\right.
\end{equation}

\subsection{Numerical Implementation}
\subsubsection{Variational Formulation}
We numerically solve \eqref{eq:de} and \eqref{eq:bc} via the finite element method 
and we briefly summarize the key steps to achieve the variational formulation.
First, multiply \eqref{eq:de} with a test function $v(\mathbf{r})$ from a suitable function space and integrate over $\Omega$, obtaining

\begin{equation}
    \label{eq:de-multiplied-with-v}
    \int_{\Omega}\mu_a(\mathbf{r})\Phi(\mathbf{r}) v(\mathbf{r}) \d\mathbf{r}-\int_{\Omega} \nabla\cdot (D(\mathbf{r})\nabla\Phi(\mathbf{r}))v(\mathbf{r})\d\mathbf{r} =\int_{\Omega}S(\mathbf{r})v(\mathbf{r})\d \mathbf{r}.
\end{equation}

Then use integration by parts to get

\begin{equation}
    \label{eq:integration-by-parts}
    \int_{\Omega}\mu_a(\mathbf{r})\Phi(\mathbf{r}) v(\mathbf{r}) \d\mathbf{r}+\int_{\Omega} D(\mathbf{r})\nabla\Phi(\mathbf{r})\cdot\nabla v(\mathbf{r})\d\mathbf{r}-\int_{\d\Omega}(D(\mathbf{\r})\underset{\frac{\Phi(\mathbf{r})}{2D(\mathbf{r})}}{\underbrace{\nabla\Phi(\mathbf{r})\cdot\mathbf{n}})}v(\mathbf{r})\d s =\int_{\Omega}S(\mathbf{r})v(\mathbf{r})\d \mathbf{r}.
\end{equation}

Reorganize the terms and conclude with

\begin{equation}
    \label{eq:final-pde}
    \int_{\Omega}(\left(\mu_a(\mathbf{r})\Phi(\mathbf{r}) - S(\mathbf{r})\right) v(\mathbf{r})+ D(\mathbf{r})\nabla\Phi(\mathbf{r})\cdot\nabla v(\mathbf{r}))\d\mathbf{r} = \int_{\d\Omega}\frac{\Phi(\mathbf{r})v(\mathbf{r})}{2}\d s.
\end{equation}

\subsubsection{Meshes}
\begin{figure}[tb]
    \centering
    \includegraphics[width=0.8\linewidth]{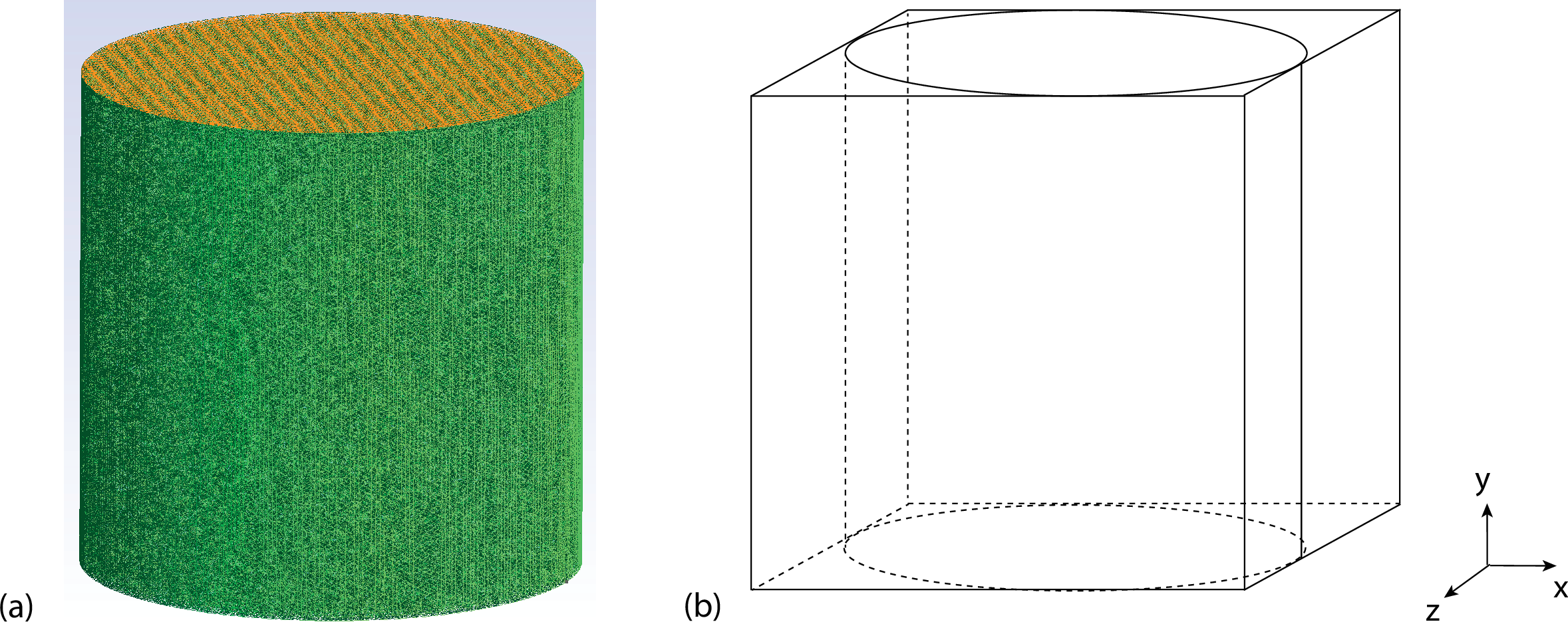}
    \caption{(a) Finite-element mesh of a cylinder generated using Gmsh.
    (b) Relation between the grid and the finite-element mesh.}
    \label{fig:relation-meshes}
\end{figure}

We represent the domain $\Omega$ with two types of meshes of uniform mesh elements.
The first type is a grid composed of boxes with size determined by the physical size $(Z, X, Y)$ of $\Omega$ and the number of sampling points $(N_{\text{z}}, N_{\text{x}}, N_{\text{y}})$ in each dimension: $(Z/N_{\text{z}}, X/N_{\text{x}}, Y/N_{\text{y}})$.
This type is used in the construction of the forward model and the unmixing algorithm (due to matrix operations).
The second type is a finite element mesh that discretizes the cylinder domain directly into tetrahedrons with mesh size $\Delta s$, which we choose to be $\max(Z/N_{\text{z}}, X/N_{\text{x}}, Y/N_{\text{y}})/2$ to reach a balance between computational accuracy and speed (see Fig. \ref{fig:relation-meshes} (a)).
This is done using Gmsh, an open-source 3D finite element mesh generator \cite{geuzaine_gmsh_2009}.
This type of mesh is used to solve the diffusion equation.
For computational speed, we precompute and save the 3D mesh with defined physical sizes and only load it when needed.

The relation between these two meshes is illustrated in Fig. \ref{fig:relation-meshes} (b).
To be precise, the box mesh tightly fits the cylinder.
The conversion between the two meshes is necessary when we compute the fluence using optical parameter maps defined on a grid and when we cast the solution of the fluence from the finite element mesh to the grid.
We use linear interpolation in both cases.

\subsubsection{Solving the Equation}
\eqref{eq:final-pde} is solved using Fenicsx, an open-source library for the numerical solution of partial differential equations \cite{LoggEtal_11_2012, AlnaesEtal2015}.
We provide a summary in Algorithm \ref{alg:fenics}.
\begin{algorithm}[tb]
\caption{Algorithm to solve the diffusion equation using Fenicsx}
\label{alg:fenics}
    \begin{algorithmic}[1]
        \State \textbf{Input} $\mu_a(\mathbf{r}), \mu_s^{'}(\mathbf{r})$, and $S(\mathbf{r})$ as 3D arrays
        \State Define mesh and function space $V$ of type ``continuous Galerkin'' of order 1
        \State Convert $\mu_a(\mathbf{r}), \mu_s^{'}(\mathbf{r})$, and $S(\mathbf{r})$ to functions in $V$
        \State Assemble \eqref{eq:final-pde} into a linear form $a(\Phi, v) = L(v)$
        \State Solve the linear form to get solution $\phi_h$
        \State Convert the finite-element solution $\phi_h$ over a cylinder to a 3D box array $\phi^{3\text{D}}$ that contains $\phi_h$
        \State Slice a cross-section of $\phi^{3\text{D}}$ to obtain a 2D fluence map $\phi^{2\text{D}}$
        \State \textbf{Output} 2D array $\phi^{2\text{D}}$
    \end{algorithmic}
\end{algorithm}

Note that the finite-element solution $\phi_h$ is over a cylindrical domain.
We convert it to a 3D array that tightly contains the cylinder.
The output of the fluence computation is a 2D cross-sectional image of the original 3D map, which we use to construct the forward model as described in Fig. 2 of the manuscript.

\section{Effectiveness of the Regularization}\label{sec:regularization}
\begin{figure}[h!]
    \centering
    \includegraphics[width=0.8\textwidth]{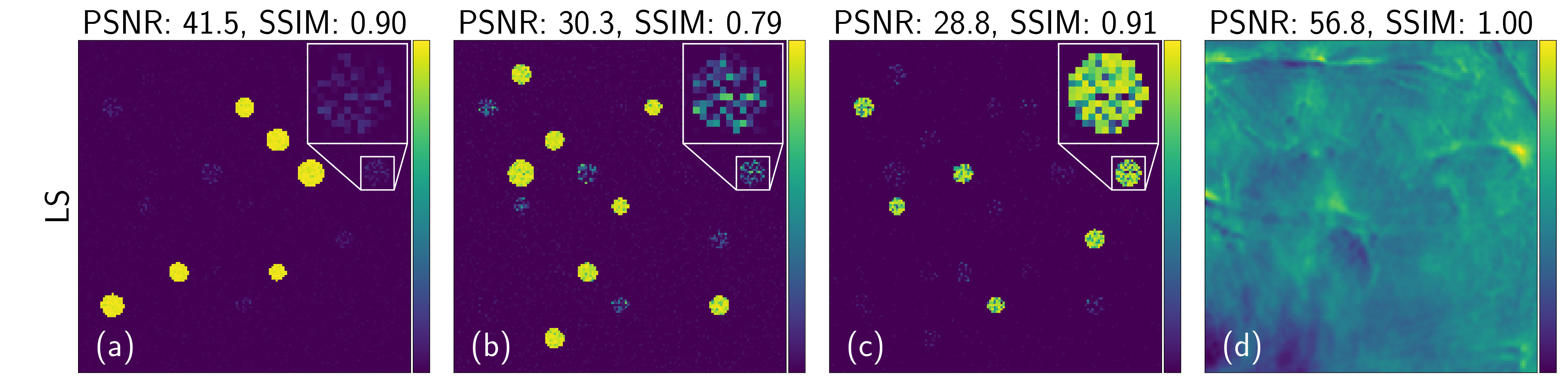}
    \includegraphics[width=0.8\textwidth]{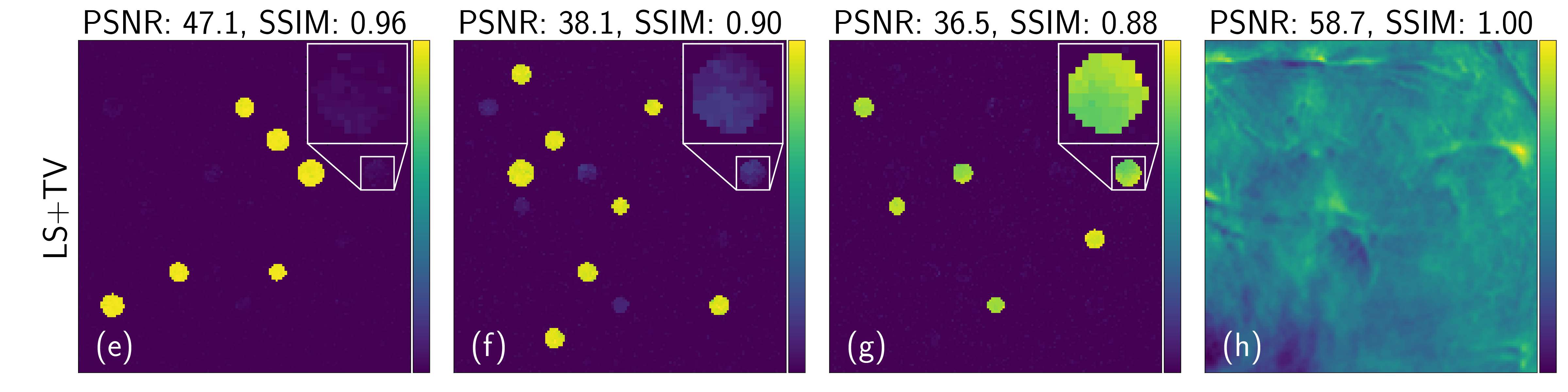}
    \includegraphics[width=0.8\textwidth]{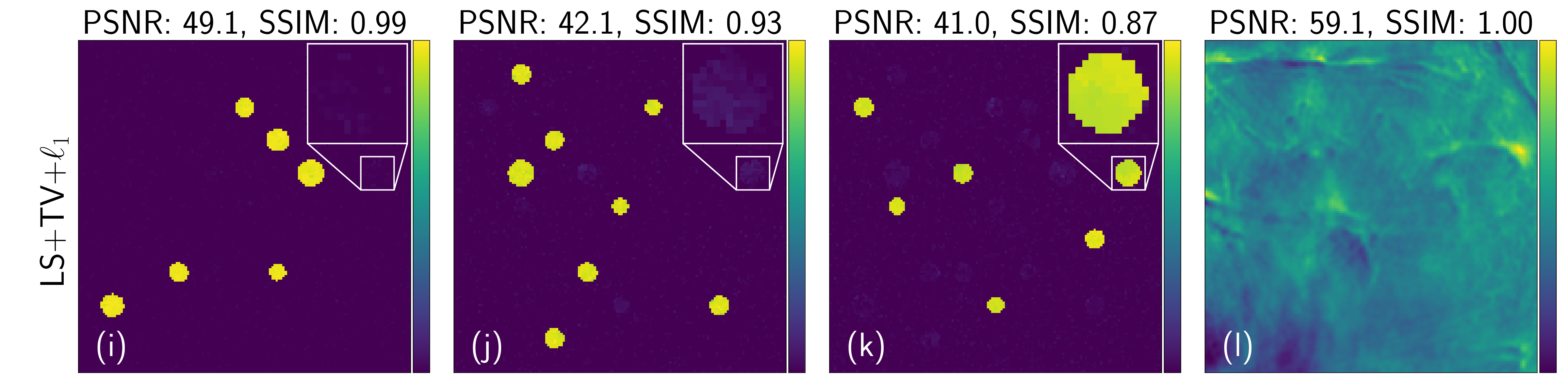}\\
    \includegraphics[width=0.8\textwidth]{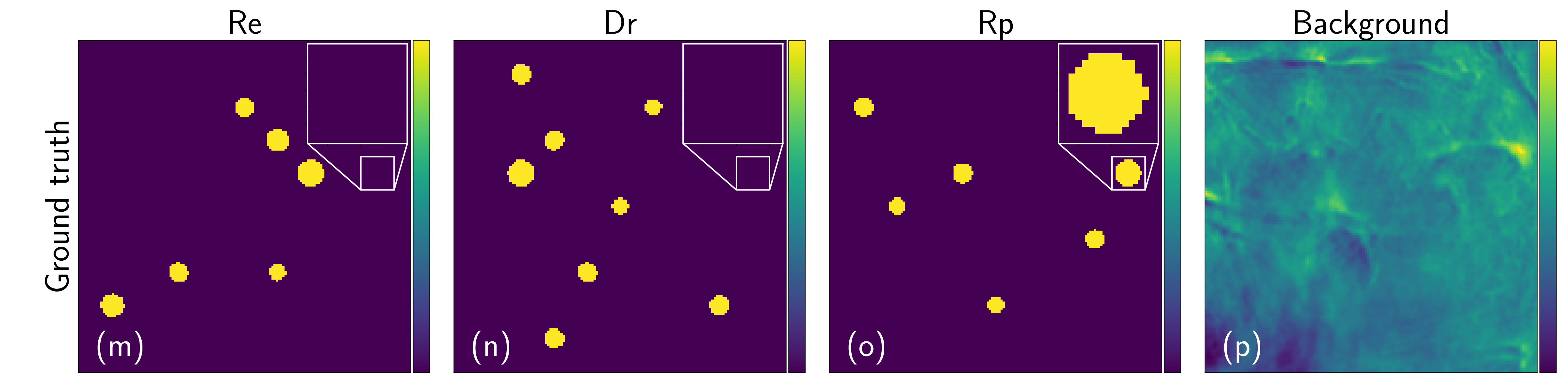}
    \includegraphics[width=0.8\textwidth]{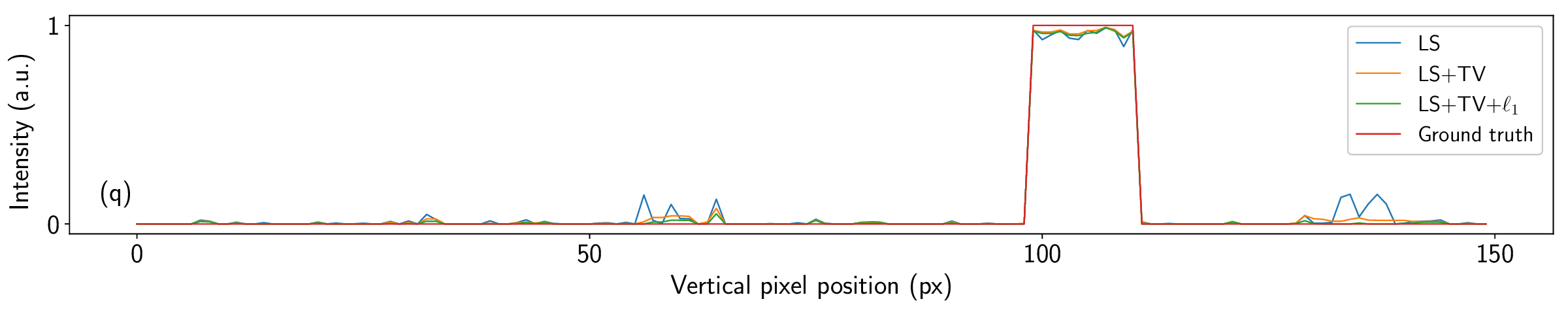}
    \includegraphics[width=0.8\textwidth]{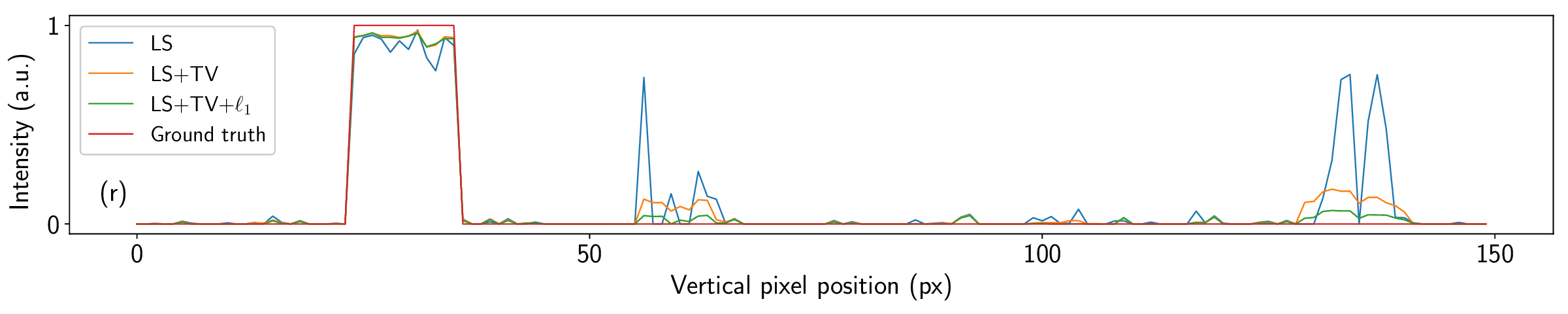}
    \includegraphics[width=0.8\textwidth]{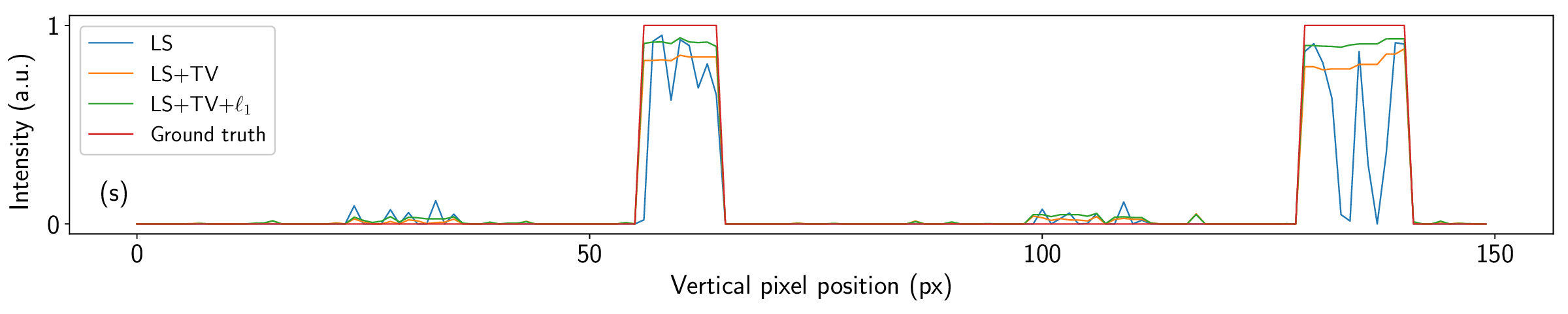}
    \caption{Unmixing results of a simulated phantom of beads mimicking the experimental phantom.
    The first three rows each contains reconstructed distribution maps of the three rsOAPs shown in column one to three and the non-switching background image in the last column. 
    The PSNR and SSIM of each reconstruction is indicated on top of each image.
    Row one (a)-(d): least-square solution.
    Row two (e)-(h): least-square with total variation regularization.
    Row three (i)-(l): least-square with total variation and $\ell_1$ regularization.
    Row four (m)-(p): ground truth.
    (q)-(s): horizontal intensity profiles at the 60th row of pixels of the reporter Re (q), Dr (r), and Rp (s) for each reconstruction method.
    The intensities of all images are normalized to $[0, 1]$ for visual comparison.}
    \label{fig:simulation}
\end{figure}
We show the effectiveness of the proposed regularization in our unmixing framework with the following simulation.
We design a 2D numerical phantom of size $(150\times150)$px in which we randomly insert three groups of disk targets of diameter between $7$px and $12$px that represent the photo-switching reporters on a heterogeneous background.
The background image is taken from an image (cropped to contain only areas of the sample) of the last pulse of the first OFF-switching cycle from the experimental dataset of the mouse model.
Each target object has homogeneous intensity.
The ratio of the initial intensities between the three groups and the maximum intensity of the background is 2:1:1:5, meaning that the background signal is stronger than that of the reporters (which mimics the scenario of a blood-rich sample).
Each group has the same kinetic constant as one of the three rsOAPs used in the experiment.
The forward model is used to generate measurements of a set of a total of 100 2D OA switching images of size $(150\times150)$px.
A level of $1\%$ Gaussian white noise is added to the simulated measurements to mimic to the noise level in the experimental datasets (see Fig. \ref{fig:noise-compare} (a)).

We tested three reconstruction approaches, namely, least-square only (LS), least-square regularized with total variation on the spatial maps of each reporter (LS+TV), and least-square regularized with TV and additionally $\ell_1$ among the reporters (LS+TV+$\ell_1$).
We compare in Fig. \ref{fig:simulation} the results of using these methods against the ground truth.
In all the scenarios, the three groups of reporters are successfully extracted from the non-switching background which is well recovered with a perfect structural similarity index (SSIM) of 1.0, though the peak signal-to-noise ratio (PSNR) varies among different methods.

Without any regularization, the unmixing performs well only on the first group of reporters, which has the largest kinetic constant, achieving an SSIM of 0.9 out of 1.
However, the algorithm struggles to separate the second and the third groups.
This is indicated by the presence of 
reporters belonging to the third group appearing in the map of the second, leading to an SSIM of only 0.79 for the second group.

In fact, reporters of the third group ``bleed'' into the maps of the first two groups (see the zoomed-in area in Fig.\ref{fig:simulation} (a) and (b)).
After having added a TV regularizer for each group and tuned the regularization weights, we observe in Fig. \ref{fig:simulation} (e)-(g) that TV not only reduces noise inside the beads, giving a much smoother intensity distribution for each of them, but also greatly mitigates the bleed-through issue if we compare Fig. \ref{fig:simulation} (b) and (f). 
The PSNR of each group enjoys an average boost of 7dB due to the noise reduction and the SSIM of reporters Re and Dr is improved by 7\% on average.
However, there is still a hint of a remaining bleed-through in Fig. \ref{fig:simulation} (f), which is solved by the further addition of an $\ell_1$ regularizer to enforce sparsity at each pixel. 
This proves to effectively solve the bleed-through problem and the intensity of each bead is more uniform, increasing the PSNR of each map by another 2-5dB (see Fig. \ref{fig:simulation} (i)-(k)).

To gain more insight on the reconstruction quality, we compare the intensity profiles along a representative horizontal line at the 60th row in Fig. \ref{fig:simulation} (q)-(s). 
We see that LS+TV+$\ell_1$ outperforms LS only and LS+TV in the sense that, on one hand, the signal is correctly recovered at nearly its 100\% intensity with very little fluctuation inside the bead (see signals at location 55px and 140px in Fig. \ref{fig:simulation} (s) for a good example) and that,
on the other hand, LS+TV+$\ell_1$ does not suffer from the bleed-through issue on locations where there should not be any signal compared to the other two approaches.
Taking the signals at location 55px and 140px in Fig. \ref{fig:simulation} (q) and (r) as an example, LS only and LS+TV show relatively prominent peaks at these two locations where the signals from the third group should not appear at all, while LS+TV+$\ell_1$ shows only flat tiny bumps. Hence, we conclude that our proposed reconstruction method, which combines TV and $\ell_1$ regularization with least square, yields the best reconstruction results.

\section{Estimation of the kinetic constants}\label{sec:kinetics}
\begin{figure}[h]
    \centering
    \includegraphics[width=0.4\textwidth]{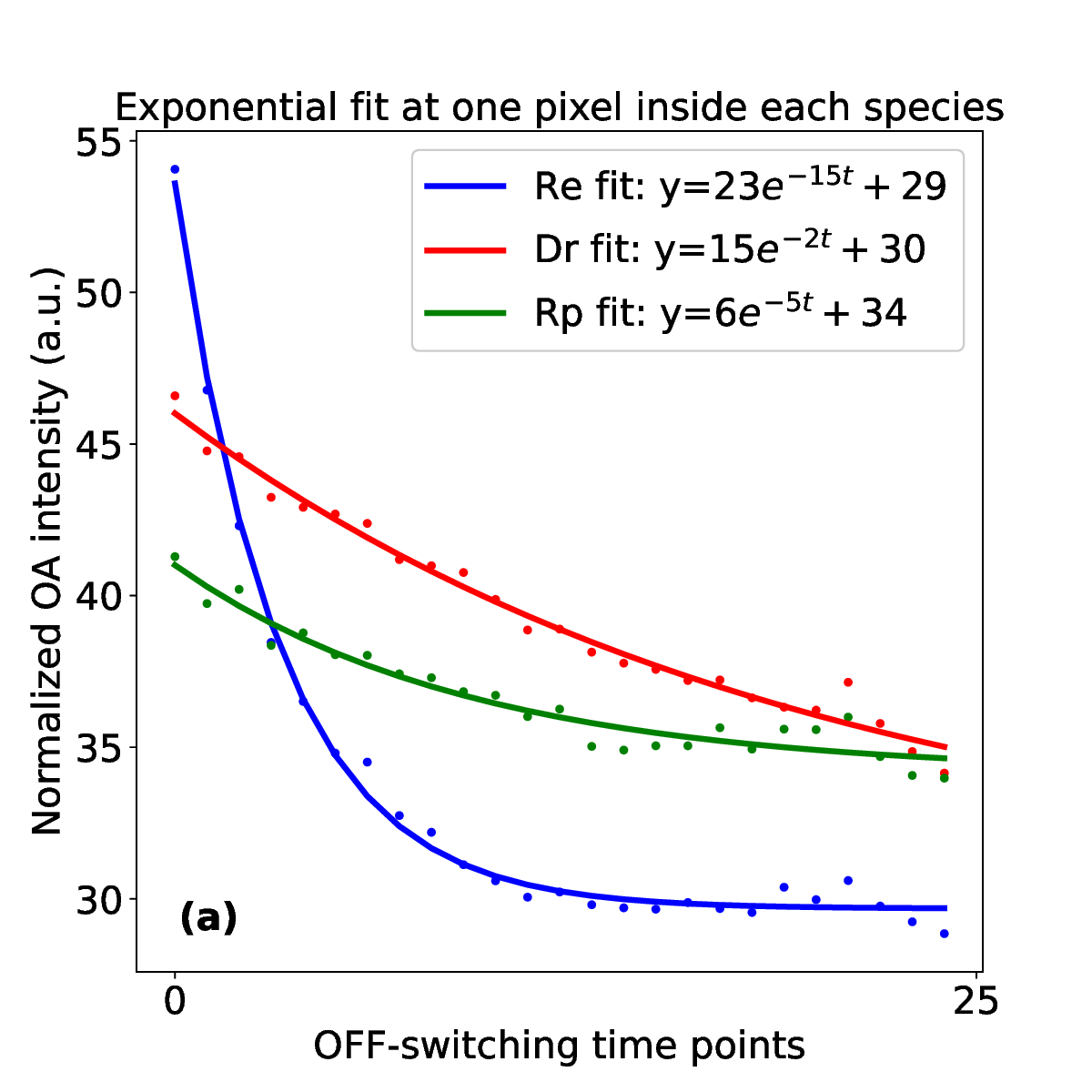}
    \includegraphics[width=0.4\textwidth]{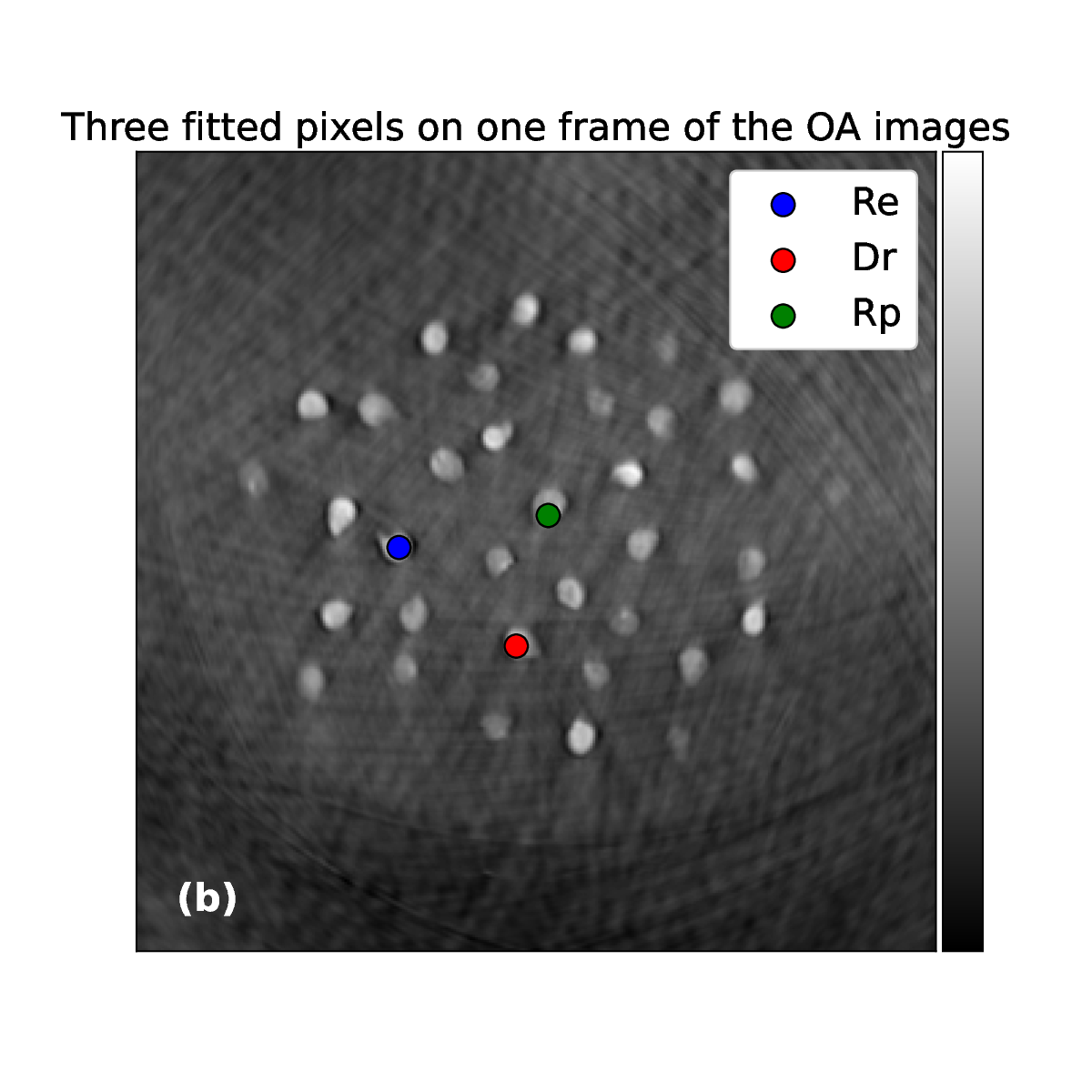}
    \includegraphics[width=0.8\textwidth]{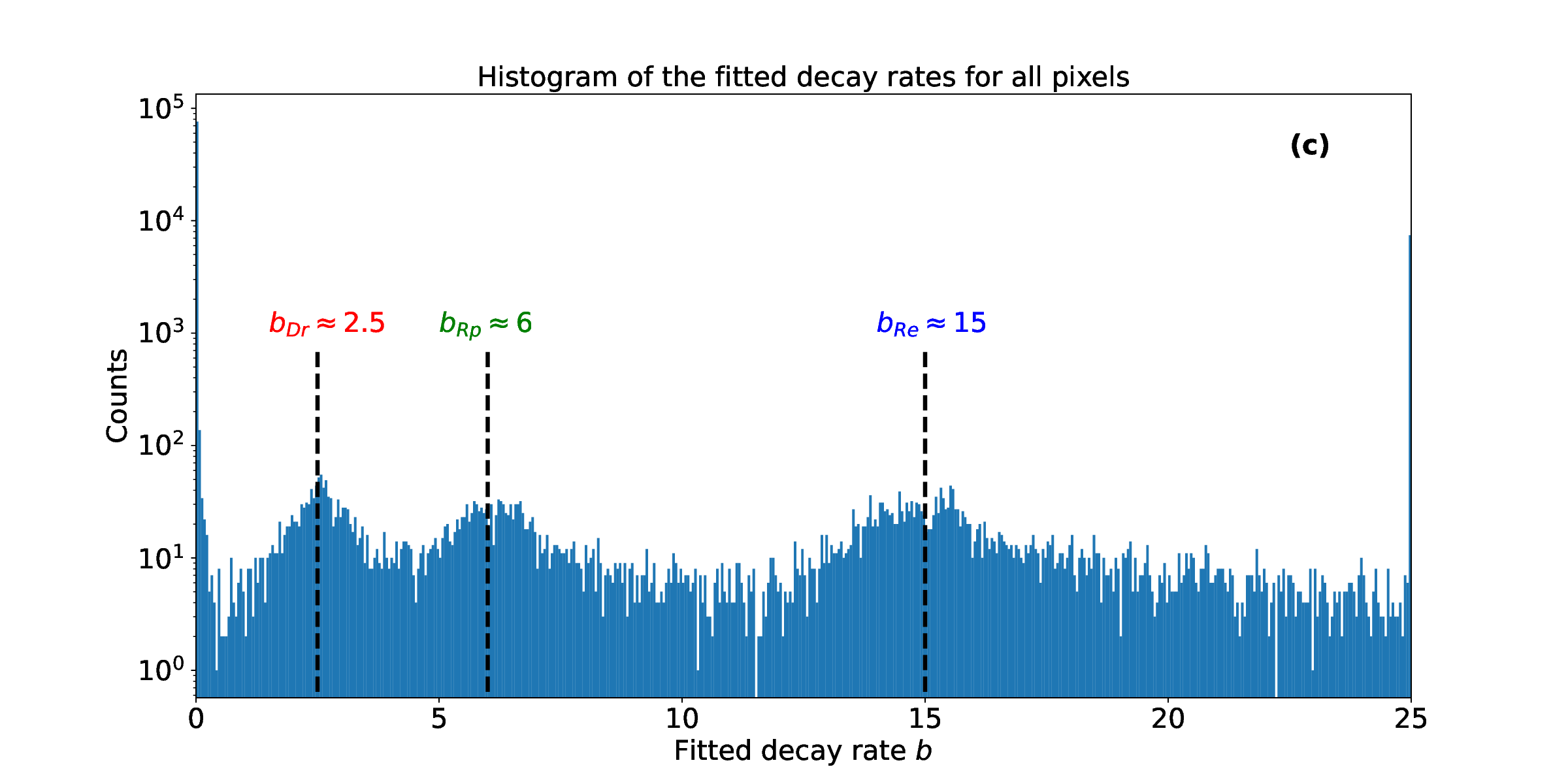}
    \caption{Estimation of the decay rates for the three species based on the OFF-switching dataset of the experimental phantom of beads.
    (a) Fitting results at three representative pixels, each inside one of the three rsOAPs. 
    The original data is the scatter points, the curves are the fitted results.
    (b) Location of the three pixels used in (a). The OA image is the first frame of the OFF-switching series.
    (c) Histogram (partial) of the fitted decay rate $b$ for all the pixels in the OA image. 
    Based on the fitting result in (a), only the part of the histogram where the values within the range (<25) is displayed.
    }
    \label{fig:find-k}
\end{figure}

The kinetic constants $\bar{k}$ reported in Table \ref{tab:reporters} are measured experimentally and depend on the pulse energy and the subsequent unknown light fluence $\Phi_m$ during the measurement via $k = \frac{\bar{k}}{\Phi_m}$, where $k$ is the intrinsic constant.
To estimate $\Phi_m$, we first use an empirical method to retrieve the decay rates $\{b_{\text{Re}}, b_{\text{Dr}}, b_{\text{Rp}}\}$ of each species from the experimental dataset.
Then, we determine $\Phi_m$ from 
\begin{equation}
    \Phi_m = \sqrt[3]{\frac{\bar{k}_{\text{Re}}}{b_{\text{Re}}}\frac{\bar{k}_{\text{Dr}}}{b_{\text{Dr}}}\frac{\bar{k}_{\text{Rp}}}{b_{\text{Rp}}}},
\end{equation}\label{eq:average}
which is a cubic square root of the estimated results from all three species. 
To estimate the $\{b_{\text{Re}}, b_{\text{Dr}}, b_{\text{Rp}}\}$, we fit an exponential function 
\begin{equation}
    y = a\e^{-bt} + c
\end{equation}\label{eq:exp-fit}
at each pixel to obtain a map of the decay rates $b$.
Then, we approximate the OFF-switching rate of each reporter by locating the peaks in the histogram of the map of $b$. 
In Fig. \ref{fig:find-k}(a), we show an example of the fitting result at one location inside each of the three rsOAPs marked by the three dots in Fig. \ref{fig:find-k}(b) for the dataset of the experimental phantom of beads.
In Fig. \ref{fig:find-k}(c), we show the result of the empirical estimate of the kinetic constants.
Three prominent peaks are observed, indicating the decay rates of the three species.
We take a value around these peaks as the empirical estimate for $\{b_{\text{Re}}, b_{\text{Dr}}$ and $ b_{\text{Rp}}\}$.
Plugging the estimated $\{b_{\text{Re}}, b_{\text{Dr}}, b_{\text{Rp}}\}$ in Fig. \ref{fig:find-k} and {$\bar{k}_{\text{Re}}, \bar{k}_{\text{Dr}},\bar{k}_{\text{Rp}}$} in Table \ref{tab:reporters} into \eqref{eq:average}, we obtain that $\Phi_m = 0.3$.

\end{document}